\documentclass[preprint,showpacs,preprintnumbers,amsmath,amssymb]{revtex4}

\usepackage{graphicx}
\usepackage{dcolumn}
\usepackage{bm}

\begin{document}


\title{Leading Order Calculation of Shear Viscosity \\ in Hot Quantum Electrodynamics from Diagrammatic Methods}


\author{Jean-S\'{e}bastien Gagnon}
\email{gagnonjs@physics.mcgill.ca}

\author{Sangyong Jeon}%
\email{jeon@physics.mcgill.ca}

\affiliation{Physics Department, McGill University, 3600 University street, Montr\'{e}al, Canada, H3A 2T8}

\date{\today}

\begin{abstract}
We compute the shear viscosity at leading order in hot Quantum Electrodynamics.  Starting from the Kubo relation for shear viscosity, we use diagrammatic methods to write down the appropriate integral equations for bosonic and fermionic effective vertices.  We also show how Ward identities can be used to put constraints on these integral equations.  One of our main results is an equation relating the kernels of the integral equations with functional derivatives of the full self-energy; it is similar to what is obtained with two-particle-irreducible effective action methods.  However, since we use Ward identities as our starting point, gauge invariance is preserved.  Using these constraints obtained from Ward identities and also power counting arguments, we select the necessary diagrams that must be resummed at leading order.  This includes all non-collinear (corresponding to 2 to 2 scatterings) and collinear (corresponding to 1+N to 2+N collinear scatterings) rungs responsible for the Landau-Pomeranchuk-Migdal effect.  We also show the equivalence between our integral equations obtained from quantum field theory and the linearized Boltzmann equations of Arnold, Moore and Yaffe obtained using effective kinetic theory.
\end{abstract}

\pacs{Valid PACS appear here}
\maketitle






\section{Introduction}
\label{sec:Intro}

Shear viscosity is a transport coefficient that characterizes the diffusion of momentum transverse to the direction of propagation.  It has attracted a lot of attention in the realm of relativistic heavy ion collisions lately since it could be an important parameter in Quark-Gluon Plasma (QGP) evolution.  Its value is still a matter of debate in the heavy ions community (see for example \cite{PHENIX_2005}).  On the theoretical side, computations in strongly coupled Super Yang-Mills theories \cite{Policastro_etal_2001,Kovtun_etal_2005} indicate a low viscosity while kinetic theory calculations \cite{AMY_2003b} indicate a higher viscosity near the critical temperature in the weak coupling limit.  See also Ref.~\cite{Huot_etal_2006} for a comparison of the shear viscosity to entropy ratio in Super Yang-Mills and Quantum Chromodynamics (QCD) at weak coupling, showing a significant difference between the two near the critical temperature.  Recent lattice calculations in pure SU(3) \cite{Nakamura_Sakai_2005,Meyer_2007} show that the shear viscosity is very close to the conjectured Kovtun-Son-Starinet bound \cite{Kovtun_etal_2005}; however, these studies are not complete since they need some external input to solve the ill-posed inverse problem of obtaining the (continuous) spectral density from the Euclidean correlator (computed at a discrete set of points).  On the phenomenological side, elliptic flow data (see for example \cite{Lacey_Taranenko_2006,Teaney_2003}) and transverse momentum correlations \cite{Gavin_etal_2006} point toward a low viscosity (see also \cite{Molnar_Huovinen_2005} for another interpretation) while others argue that color glass condensate initial conditions \cite{Hirano_etal_2006} and plasma instabilities \cite{Asakawa_etal_2006} could explain the data without invoking a necessarily low viscosity.  Viscous hydrodynamic simulations are on the way \cite{Chaudhuri_2006,Heinz_etal_2006,Muronga_Rischke_2004,Muronga_2004} and will possibly shed some light on these issues.  From the above, it is clear that understanding the inner workings of shear viscosity (especially in gauge theories) is important. 

In the present paper, we are interested in computing the shear viscosity at leading order in gauge theories from quantum field theory, starting from the Kubo relation and using diagrammatic methods.  This calculation has been in done in scalar theory \cite{Jeon_1995} and has been reproduced since then using different methods (e.g. real-time formalism \cite{Wang_Heinz_1999,Wang_Heinz_2003,Carrington_etal_2000}, direct ladder summation in Euclidean space \cite{Basagoiti_2002}, 2PI effective action methods \cite{Aarts_Martinez_2003} and quantum kinetic field theory derived from the closed-time-path 2PI effective action \cite{Calzetta_etal_2000}).  The difficulty of the calculation lies in the fact that, due to ``pinch'' singularities, an infinite number of ladder diagrams must be resummed to obtain the result even at leading order.  The calculation is even more involved in gauge theories; in addition to pinch singularities, the appearance of collinear singularities makes the resummation of another class of ladder diagrams necessary \cite{AMY_2003b,AMY_2003a,Gagnon_Jeon_2007}. There exists some attempts at computing the shear viscosity in gauge theories from quantum field theory (e.g. real-time formalism \cite{Defu_2005}, direct ladder summation in Euclidean space \cite{Basagoiti_2002}, 2PI effective action methods \cite{Aarts_Martinez_2005}), but to the best of the authors' knowledge, none of these approaches go beyond leading log order accuracy (i.e. with corrections suppressed by $O(g\ln(g^{-1}))$) or the large $N_{f}$ approximation.

A more convenient way to compute transport coefficients is to use kinetic theory.  The leading order shear viscosity in hot gauge theories has been computed by Arnold, Moore and Yaffe (AMY) using an effective kinetic theory where zero temperature masses are replaced by thermal masses \cite{AMY_2003b,AMY_2003a}.  This calculation consistently includes the physics of pinch singularities, collinear singularities and the Landau-Pomeranchuk-Migdal (LPM) effect.  Contrary to the scalar case where the equivalence between the quantum field theory approach and the kinetic theory approach has been shown \cite{Jeon_1995,Jeon_Yaffe_1996}, the landmark calculation of AMY has never been verified using quantum field theory.  Since the perturbative results of AMY for the shear viscosity is often quoted in the RHIC community, it would be good to have a first principle quantum field theory proof of this kinetic theory calculation.

It is thus the goal of this paper to compute the shear viscosity at leading order in hot Quantum Electrodynamics (QED) from purely diagrammatic methods and show the equivalence with the kinetic theory results of AMY.  A similar calculation for electric conductivity has been performed in \cite{Gagnon_Jeon_2007,Gagnon_Jeon_2007_erratum}.  In the following, we use the same general method outlined in Ref.~\cite{Gagnon_Jeon_2007,Gagnon_Jeon_2007_erratum}, emphasizing the differences with the electric conductivity calculation.  The rest of the paper is organized as follows.  Section~\ref{sec:Background} presents our notation and some background material on transport coefficients, both in scalar and gauge theories.  We present our constraint on the ladder kernels coming from Ward-like identities in Sect.~\ref{sec:Ward_identity}.  Power counting arguments are shown in Sect.~\ref{sec:Power_counting}, in order to determine which rungs should be kept in the resummation.  The final expressions for shear viscosity, including collinear physics and the LPM effect, are presented in Sect.~\ref{sec:Integral_equations}, where we also show the equivalence with the kinetic theory results of AMY.  We finally conlude in Sect.~\ref{sec:Conclusion}.

\section{Background Material}
\label{sec:Background}

\subsection{Notation and definitions}
\label{sec:Notation}

We present here a summary of our notation along with a list of the various finite temperature field theory quantities that we use throughout the paper.  Latin indices run from 1 to 3 and represent space components while Greek indices run from 0 to 3 and represent spacetime components.  Boldface, normal and capital letters denote 3-momenta, 4-momenta and Euclidean 4-momenta, respectively.  We use the metric convention $\eta_{\mu\nu} = (1,-1,-1,-1)$.  The subscripts $B$, $F$ attached to a quantity refer to its bosonic or fermionic nature (except for self-energies and widths, where we use special notations).  The subscripts $R$, $I$ mean real or imaginary part and the superscripts $\rm ret$, $\rm adv$, $\rm cor$ mean retarded, advanced or autocorrelation (i.e. average value of the anti-commutator).  A bar over a quantity means that the gamma matrix structure is explicitly taken out (e.g. $G^{\mu}(k) \equiv \gamma^{\mu}\bar{G}(k)$).

Spectral densities are important quantities (especially at finite temperature).  Any 2-point function can be expressed in terms of them and they satisfy a number of general properties (see for example \cite{LeBellac_2000,Forster_1975}).  Spectral densities can be expressed in terms of commutators of fields (see for example \cite{Chou_etal_1985,Fetter_Walecka_2003}), but it is sufficient for our purposes to give their explicit expressions in momentum space.  Free field spectral densities are given by \cite{Jeon_Ellis_1998}:
\begin{eqnarray}
\label{eq:free_spectral_density_boson}
\rho_{B}(k) & = & \mbox{sgn}(k^{0})2\pi\delta((k^{0})^{2}-E_{k}^{2}) \\
\label{eq:free_spectral_density_fermion}
\rho_{F}(k) & = & \left[2\pi\delta(k^{0} - E_{k})h_{+}(\hat{k}) + 2\pi\delta(k^{0} + E_{k})h_{-}(\hat{k})\right]
\end{eqnarray}
where $\mbox{sgn}(k^{0})$ is the sign function, $E_{k} \equiv |\mathbf{k}|$, $h_{\pm}(\hat{k}) \equiv (\gamma^{0} \mp \mathbf{\gamma}\cdot\hat{k})/2$ and $\hat{k} \equiv \mathbf{k}/|\mathbf{k}|$.  Note that since we consider systems where the temperature is much larger than any other scale, we put $m = 0$ in the above and all subsequent expressions when the momentum of the excitation is hard.  From the above expressions (or more generally from CPT invariance), it can be shown that the spectral densities satisfy $\rho_{B}(-k^{0}) = -\rho_{B}(k^{0})$ and $\rho_{F}(-k) = \rho_{F}(k)$ (in the massless limit).  The delta functions in Eqs.~(\ref{eq:free_spectral_density_boson})-(\ref{eq:free_spectral_density_fermion}) means that the excitations have sharply peaked energies and an infinite lifetime.  But at finite temperature, any excitation propagating in a medium has a finite lifetime due to numerous collisions with on-shell thermal excitations.  The effect of this finite lifetime is to turn the delta functions in Eqs.~(\ref{eq:free_spectral_density_boson})-(\ref{eq:free_spectral_density_fermion}) into Lorentzians, giving \cite{Jeon_1995,Basagoiti_2002}:
\begin{eqnarray}
\label{eq:resumed_spectral_density_boson}
\rho_{B}(k) & = & \frac{1}{2E_{k}} \left[\frac{\gamma_{k}}{(k^{0}-E_{k})^{2}+(\gamma_{k}/2)^{2}}-\frac{\gamma_{k}}{(k^{0}+E_{k})^{2}+(\gamma_{k}/2)^{2}}\right] \\
\label{eq:resumed_spectral_density_fermion}
\rho_{F}(k) & = & \left[\frac{\Gamma_{k}}{(k^{0} - E_{k})^{2} + (\Gamma_{k}/2)^{2}}h_{+}(\hat{k}) + \frac{\Gamma_{k}}{(k^{0} + E_{k})^{2} + (\Gamma_{k}/2)^{2}}h_{-}(\hat{k})\right]
\end{eqnarray}
The widths are given by $\gamma_{k} \equiv \Pi_{I}^{\rm ret}(k^{0} = E_{k})/E_{k}$ and $\Gamma_{k} \equiv \mbox{tr}\left[k\!\!\!/ \Sigma_{I}^{\rm ret}(k^{0} = E_{k}) \right]/2E_{k}$, where $\Pi(k)$ and $\Sigma(k)$ are the bosonic and fermionic self-energies, respectively.  Note that when the momentum $k$ is soft, perturbation theory must be re-organized and partial resummation of spectral densities is necessary (also called Hard Thermal Loop (HTL) resummations \cite{Braaten_Pisarski_1990,Taylor_Wong_1990,Braaten_Pisarski_1992}).  These resummations give rise to screening thermal masses and may also produce Landau damping.  In gauge theories, HTLs are also essential to obtain gauge invariant results.

The analysis in this paper relies heavily on the finite temperature cutting rules.  The building blocks of these rules are the four propagators of the closed-time-path or ``1-2'' formalism \cite{Schwinger_1961,Keldysh_1965}.  The time-ordered (or ``uncut'') propagators can be expressed in terms of the spectral densities \cite{Jeon_Ellis_1998}:
\begin{eqnarray}
\label{eq:11_propagators}
G_{B/F}(k) & = & i\int\frac{d\omega}{(2\pi)}\; \rho_{B/F}(\omega)\left(\frac{1\pm n_{B/F}(\omega)}{k^{0}-\omega+i\epsilon} \pm \frac{n_{B/F}(\omega)}{k^{0}-\omega-i\epsilon} \right)
\end{eqnarray}
where $n_{B/F}(k^{0})$ are the usual Bose-Einstein or Fermi-Dirac distribution functions.  The anti time-ordered propagators are just the complex conjugate of the time-ordered ones.  Wightman (or ``cut'') propagators are given by \cite{Jeon_Ellis_1998}:
\begin{eqnarray}
\label{eq:12_propagators}
\Delta_{B/F}^{+}(k) & = & (1 \pm n_{B/F}(k^{0}))\rho_{B/F}(k) \\
\label{eq:21_propagators}
\Delta_{B/F}^{-}(k) & = & \pm n_{B/F}(k^{0})\rho_{B/F}(k)
\end{eqnarray}
Our notation for the propagators is different from the one of the ``1-2'' formalism; the correspondence is $G = G^{11}$, $G^{*} = G^{22}$, $\Delta^{+} = G^{12}$ and $\Delta^{-}  = G^{21}$.  Using the above propagators, we can write down the following cutting rules \cite{Kobes_Semenoff_1985,Kobes_Semenoff_1986,Jeon_Ellis_1998}:
\begin{enumerate}
	\item Draw all the cut diagrams relevant to the problem considered, where cuts separate the unshaded ({\it i.e.} ``1'') and the shaded ({\it i.e.} ``2'') regions.
	\item Use the usual Feynman rules for the unshaded region assigning $G_{B/F}(k)$ to the uncut lines.  For the shaded region, use the conjugate Feynman rules assigning $G_{B/F}^{*}(k)$ to the uncut lines.
	\item If the momentum of a cut line crosses from the unshaded to the shaded region, assign $\Delta_{B/F}^{+}(k)$.  If the momentum of a cut line crosses from the shaded to the unshaded region, assign $\Delta_{B/F}^{-}(k)$.
	\item Divide by the appropriate symmetry factor and multiply by an overall factor of $-i$.
\end{enumerate}
These rules are analogous to the zero temperature ones and reduce to them when $T = 0$.  Note that the various shadings given by the cutting rules are not all independent: they are related by unitarity (i.e. ``vanishing of all circlings'' relation \cite{Kobes_Semenoff_1985,Kobes_Semenoff_1986,Jeon_Ellis_1998}) and the Kubo-Martin-Schwinger (KMS) relations (e.g. \cite{LeBellac_2000}, see also \cite{Carrington_Heinz_1998,Carrington_etal_2000b,Wang_Heinz_2002} for 3 and 4-point functions KMS relations).

%
%

In some cases, it is convenient to work with linear combinations of shadings.  A particularly useful one is the Keldysh (or $r$,$a$) basis in which we can write the physical functions (see for example \cite{Chou_etal_1985}):
\begin{eqnarray}
\label{eq:physical_functions}
iG_{B/F}^{ra} \;\equiv\; iG_{B/F}^{\rm ret}(k) & = & G_{B/F}(k) - \Delta_{B/F}^{-}(k) \\
iG_{B/F}^{ar} \;\equiv\; iG_{B/F}^{\rm adv}(k) & = & G_{B/F}(k) - \Delta_{B/F}^{+}(k) \\
iG_{B/F}^{rr} \;\equiv\; iG_{B/F}^{\rm cor}(k) & = & \Delta_{B/F}^{+}(k) + \Delta_{B/F}^{-}(k)
\end{eqnarray}
The $G_{B/F}^{aa}$ is identically zero in the Keldysh basis (due to unitarity).  Note that $G_{ra}$ and $G_{ar}$ do not depend explicitly on distribution functions.  Using the equilibrium expressions for the cut propagators~(\ref{eq:12_propagators})-(\ref{eq:21_propagators}), we see that the $rr$ propagator satisfies the fluctuation-dissipation theorem $iG_{B/F}^{rr}(k) = (1 \pm 2n_{B/F}(k^{0}))\rho_{B/F}(k) = (1 \pm 2n_{B/F}(k^{0}))\left(iG_{B/F}^{ra}(k)-iG_{B/F}^{ar}(k)\right)$ (see for example \cite{Forster_1975,Wang_Heinz_2002}).  Note also that any vertex in this basis must involve an odd number of $a$'s (see for example \cite{Gelis_1997}), independent of the form of the interaction.  To compute a certain diagram (topology) in the Keldysh basis, we add all possible $r$,$a$ configurations consistent with the above two constraints.

\subsection{Shear viscosity from quantum field theory}
\label{sec:Transport_QFT}

Shear viscosity is a transport coefficient that characterizes the diffusion of transverse momentum due to collisions in a medium.  It is roughly proportional to the mean free path of excitations in the medium; simple parametric estimates using kinetic theory shows that the shear viscosity in hot QED (i.e. where hot means that the temperature is much larger than the electron mass) behaves as (to leading-log accuracy):
\begin{eqnarray}
\label{eq:Parametric_estimate}
\eta & = & C\frac{T^{3}}{e^{4}\ln e^{-1}}
\end{eqnarray}
where $T$ is the temperature of the medium, $e$ is the electromagnetic coupling constant and $C$ is a numerical coefficient that can only be obtained from a detailed analysis \cite{AMY_2003b,AMY_2003a,AMY_2000}.  The coupling constant dependence of Eq.~(\ref{eq:Parametric_estimate}) is expected, since transport coefficients are roughly proportional to the mean free path and thus inversely proportional to the scattering cross section of the processes responsible for transport (Coulomb scattering in the present case).

From linear response theory and Ward identities, it is possible to express transport coefficients in terms long distance correlators between conserved currents.  The resulting Kubo formula for the shear viscosity $\eta$ is given by \cite{Hosoya_etal_1984,Jeon_1995,Kapusta_Gale_2006}:
\begin{eqnarray}
\label{eq:Kubo_relations}
\eta & = & \frac{\beta}{20} \lim_{k^{0} \rightarrow 0,\;{\bf k}=0} \int d^{4}x\; e^{i k\cdot x}\; \langle \pi_{ij}(t,{\bf x})\pi^{ij}(0) \rangle_{\rm eq}
\end{eqnarray}
where $\pi_{ij}(x) = T_{ij}(x) - \delta_{ij}T_{i}^{i}(x)/3$ is the traceless part of the stress tensor.  Note that the averages in Eq.~(\ref{eq:Kubo_relations}) are done with respect to an equilibrium density matrix even if the shear viscosity is a non-equilibrium quantity; Eq.~(\ref{eq:Kubo_relations}) thus lends itself to a diagrammatic analysis and allows one to compute the shear viscosity from first principles.

\begin{figure}
\resizebox{3in}{!}{\includegraphics{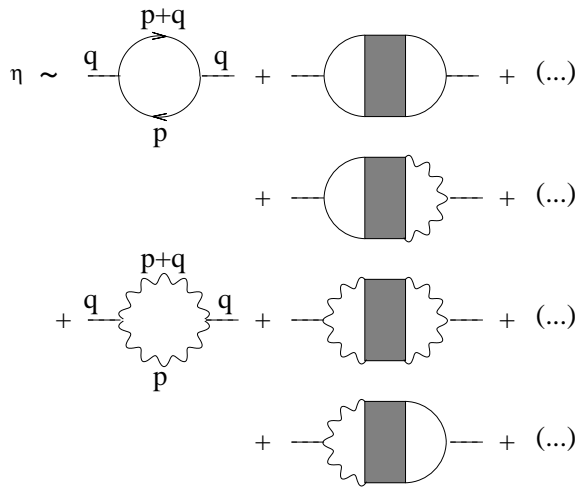}}
\caption{\label{fig:ladder_diagrams_shear} Diagrammatic expansion in ladder diagrams of the Kubo relation~(\ref{eq:Kubo_relations}).  Dotted lines represent stress insertions and grey squares represent 4-point functions (called ``rungs'').  When the external momentum $q$ goes to zero, the two ``side rail'' propagators have the same momentum and produce a ``pinch'' singularity.}
\end{figure}

In QED, the (traceless) stress tensor has the form (at leading order in the coupling) \cite{Basagoiti_2002}:
\begin{eqnarray}
\label{eq:Operator_insertion}
\pi^{ij}(k) & = & \frac{i}{2}\bar{\psi}\left[\gamma^{i}k^{j} + \gamma^{j}k^{i} - \frac{2\delta^{ij}}{3}({\bf \gamma}\cdot {\bf k})\right]\psi - A_{s}\left[k^{i}k^{j} - \frac{{\bf k}^{2}\delta^{ij}}{3}\right]\delta^{st}A_{t}
\end{eqnarray}
where $\psi$ and $A_{s}$ are electron and photon field operators, respectively.  From Eq.~(\ref{eq:Operator_insertion}), we see that such an operator insertion leads to both bosonic and fermionic vertices.  Figure~\ref{fig:ladder_diagrams_shear} shows the diagrammatic expansion of the Kubo relation~(\ref{eq:Kubo_relations}).  Naively, only one-loop diagrams should be kept at lowest order, implying no mixing between bosonic and fermionic vertices.  But these diagrams suffer from so-called ``pinch'' singularities that make their evaluation more complicated than it appears.  The argument goes as follows.  The low frequency limit in the Kubo relation~(\ref{eq:Kubo_relations}) gives rise to products of propagators $G(p)$ with the same momentum.  Due to the pole structure of finite temperature propagators (one in each quadrant), one faces situations when the integration contour is ``pinched'' between two poles on opposite sides of the real axis in the complex $p^{0}$ plane when two propagators with the same momentum are multiplied together.  In such a case, viscosity diverges as $\eta \sim \int dp^{0}\; G(p)G(p) \sim 1/\epsilon$, where $\epsilon$ is the usual infinitesimal $i\epsilon$ prescription.  These ``pinch'' singularities are of course artificial: since at finite temperature excitations always suffer collisions with thermal particles coming from the medium, they have a finite lifetime (i.e. $i\epsilon$ is effectively replaced by $i\Gamma$).  This finite lifetime regulates the pinch singularities and makes the shear viscosity proportional to $1/\Gamma \sim 1/e^{4}\ln e^{-1}$, similar to the parametric estimate~(\ref{eq:Parametric_estimate}).

\begin{figure}
\resizebox{5in}{!}{\includegraphics{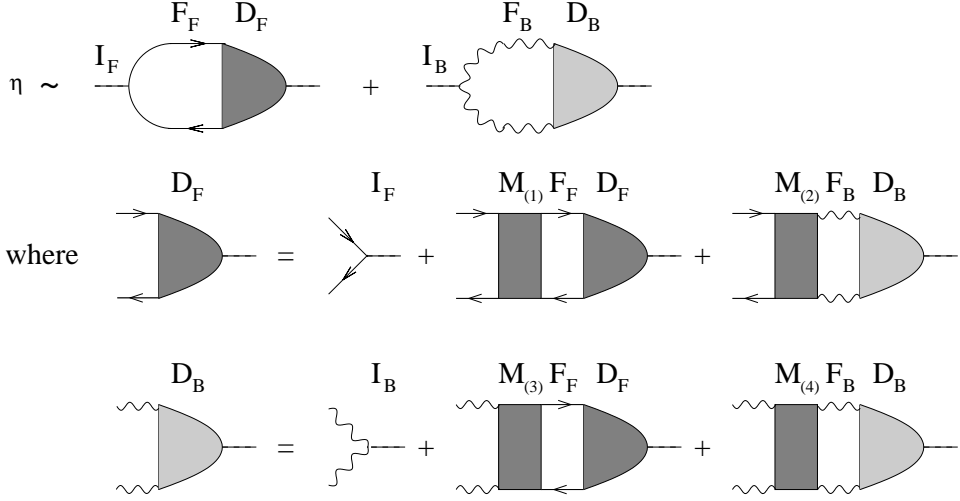}}
\caption{\label{fig:integral_equation_shear} Schematic representation of Eqs.~(\ref{eq:Kubo_relation_schematic})-(\ref{eq:integral_equation_schematic_2}).  
The symbols refer directly to the equations: ${\cal I}_{B/F}$ are stress tensor insertions, ${\cal M}_{(i)}$ are rungs, ${\cal F}_{B/F}$ represent pairs of side rail propagators and ${\cal D}_{B/F}$ are effective vertices.}
\end{figure}

The use of resummed propagators to regulate pinch singularities also makes the resummation of an infinite number of ladder diagrams necessary to obtain the correct leading order result.  One way to do this infinite resummation is to write the Kubo formula in terms of two effective vertices (bosonic and fermionic) that satisfy two coupled integral equations.  Schematically, we have:
\begin{eqnarray}
\label{eq:Kubo_relation_schematic}
\eta & = & \frac{\beta}{20}\left[ \int\frac{d^{4}k}{(2\pi)^{4}}\; {\cal I}_{F}^{*}(k){\cal F}_{F}(k){\cal D}_{F}(k) + \int\frac{d^{4}k}{(2\pi)^{4}}\; {\cal I}_{B}^{*}(k){\cal F}_{B}(k){\cal D}_{B}(k)\right]
\end{eqnarray}
with the accompanying integral equations:
\begin{eqnarray}
\label{eq:integral_equation_schematic_1}
{\cal D}_{F}(k) & = & {\cal I}_{F}(k) + \int\frac{d^{4}p}{(2\pi)^{4}}\; {\cal K}_{(1)}(k,p){\cal D}_{F}(p) + \int\frac{d^{4}p}{(2\pi)^{4}}\; {\cal K}_{(2)}(k,p){\cal D}_{B}(p) \\
\label{eq:integral_equation_schematic_2}
{\cal D}_{B}(k) & = & {\cal I}_{B}(k) + \int\frac{d^{4}p}{(2\pi)^{4}}\; {\cal K}_{(3)}(k,p){\cal D}_{F}(p) + \int\frac{d^{4}p}{(2\pi)^{4}}\; {\cal K}_{(4)}(k,p){\cal D}_{B}(p)
\end{eqnarray}
where ${\cal K}_{(i)} \equiv {\cal M}_{(i)}{\cal F}$.  See Fig.~\ref{fig:integral_equation_shear} for a graphical representation of Eqs.~(\ref{eq:Kubo_relation_schematic})-(\ref{eq:integral_equation_schematic_2}).  The symbols ${\cal F}_{B/F}$ represent pairs of bosonic/fermionic ``side rail'' propagators (note that the ``ladder'' diagrams in Fig.~\ref{fig:integral_equation_shear} are on their sides, meaning that the side rails are on the top and bottom of the diagram).  In the limit $q\rightarrow 0$, the two side rail propagators that connect ${\cal M}$ and ${\cal D}$ have the same momentum and pinch, producing a $1/\Gamma_{p}$ or a $1/\gamma_{p}$ factor.  The ``rungs'' ${\cal M}$ are amputated 4-point functions that must be of the same order as ${\cal F}^{-1}$ but otherwise arbitrary.  Note that there are four categories of rungs that allows the mixing of bosonic and fermionic effective vertices.  The bare vertices ${\cal I}_{B/F}$ represent amputated stress tensor insertions.  The effective vertices ${\cal D}_{B/F}$ encode the information about the infinite resummation of ladder diagrams.  Closing them with the appropriate bare vertex and side rails, we obtain the Kubo relation~(\ref{eq:Kubo_relation_schematic}).

The nonperturbative resummation of an infinite number of ladder diagrams due to pinch singularities is a common feature of both scalar and gauge theories.  However, as is usually the case, the gauge theory calculation contains other complications, among them HTL resummations and gauge invariance.  These issues are discussed in Ref.~\cite{Gagnon_Jeon_2007} in the case of electric conductivity in QED; we briefly come back to the issue of gauge invariance in Sect.~\ref{sec:Ward_identity}.

Another very important complication of gauge theories is the sensitivity of transport coefficients to soft and collinear physics even at leading order \cite{AMY_2003b,AMY_2003a}.  As with pinch singularities, collinear singularities generically appear when two propagators with momenta $p$ and $p+q$ that are nearly collinear (i.e. ${\bf p}\cdot{\bf q} \sim O(e^{2}T^{2})$) are multiplied together and integrated over:
\begin{eqnarray}
\label{eq:Collinear_singularities}
\int dp^{0}\; G(p)G(p+q) & \sim & \int dp^{0}\; G^{\rm ret}(p)G^{\rm adv}(p+q) \nonumber \\
                         & \sim & \int dp^{0}\; \left(\frac{1}{(p^{0}+i\epsilon)-E_{p}}\right) \left(\frac{1}{(p^{0}+q^{0}-i\epsilon)-E_{p+q}}\right) \nonumber \\
			 & \sim & \frac{1}{q^{0} + (E_{p+q}-E_{p}) - 2i\epsilon}
\end{eqnarray}
There are two ways to get a divergence in Eq.~(\ref{eq:Collinear_singularities}).  First, in the limit $q\rightarrow 0$, the expression diverges as $1/\epsilon$ and corresponds to the pinch singularity case discussed previously.  Second, when $q$ is nonzero but nearly on-shell and the angle between the quasiparticles is parametrically small (i.e. $\theta_{pq} \sim O(e)$), we have $E_{p+q} \approx E_{p}\pm |{\bf q}|$ and the expression diverges as $1/\theta_{pq}^{2}$ (or $1/\epsilon$ in the perfectly collinear case).  The regularization of this new singularity using resummed propagators introduces coupling constants in the denominator and changes the power counting dramatically.  In the scalar case where there are only pinch singularities, only one resummation of an infinite number of ladder diagrams with a {\it finite} number of rung types is necessary \cite{Jeon_1995}.  In gauge theories where two types of singularities are present (pinch and collinear), two resummations are necessary.  The one due to pinch singularities is similar to the scalar case.  The second resummation takes place inside particular types of rung where collinear singularities are present, making the number of rung types effectively {\it infinite}.  This infinite class of rung types is the manifestation of the LPM effect in our approach.  In practice, the two resummations are done using integral equations embedded in each other: the integral equations for pinch singularities are shown in Eqs.~(\ref{eq:integral_equation_schematic_1})-(\ref{eq:integral_equation_schematic_2}) and the ones for collinear singularities is hidden in the four kernels ${\cal K}_{(i)}$.  We come back to these issues and the subtleties of power counting in gauge theories in Sect.~\ref{sec:Power_counting}.





\section{Constraints on the ladder kernels from Ward-like identities}
\label{sec:Ward_identity}


Ward identities relate N-point correlation functions to a linear combination of lower-point functions and are a direct result of gauge symmetry.  In QED, Ward identities must be respected in order to preserve transversality (and thus gauge invariance).  A particular case of interest to us is the Ward identity relating a vertex to a combination of two propagators.  Since the resummation of ladder diagrams in transport coefficient calculations are written in terms of effective vertices satisfying integral equations (c.f. Eqs.~(\ref{eq:integral_equation_schematic_1})-(\ref{eq:integral_equation_schematic_2})), we expect Ward identities to put constraints on these integral equations.  It is shown in Ref.~\cite{Gagnon_Jeon_2007} how to obtain these constraints for electric conductivity in QED.  In this section, we sketch how to obtain such constraints for shear viscosity in QED; we refer the reader to Ref.~\cite{Gagnon_Jeon_2007} for details (no conceptual difficulties are added in the shear viscosity case).

In general, Kubo relations express transport coefficients in terms of different conserved current insertions.  For electric conductivity, the conserved current is the usual electric current $j^{\mu}$ and the corresponding Ward identity (relevant for transport calculations) can be derived \cite{Aarts_Martinez_2002,Gagnon_Jeon_2007}.  For shear viscosity, the conserved current is the energy-momentum tensor.  Using standard procedures (e.g. \cite{Peskin_Schroeder_1995}), we can similarly derive two Ward-like identities for a $T^{\mu\nu}$ insertion, one for each type of effective vertex.  In Euclidean space, the result is (see Fig.~\ref{fig:Ward_identity} for the momentum convention):
\begin{figure}
\resizebox{5in}{!}{\includegraphics{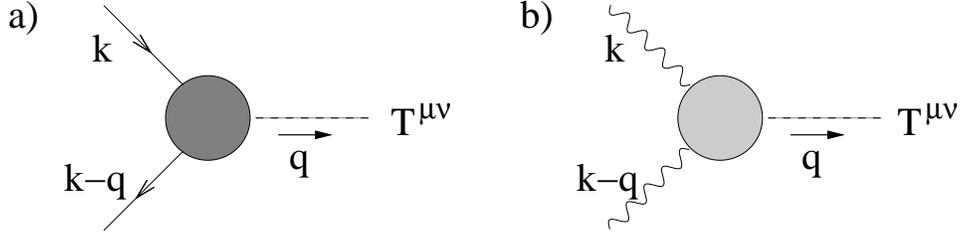}}
\caption{\label{fig:Ward_identity} Momentum convention for the Ward-like identities~(\ref{eq:Ward_energy_Euclidean_1})-(\ref{eq:Ward_energy_Euclidean_2}).  Dotted lines stand for energy-momentum tensor insertions and blobs represent the amputated effective vertices ${\cal D}$ (bosonic or fermionic).}
\end{figure}
\begin{eqnarray}
\label{eq:Ward_energy_Euclidean_1}
Q_{\nu}{\cal D}_{F}^{\mu\nu}(K,K-Q) & = & (K^{\mu}-Q^{\mu})G_{F}^{-1}(K) - K^{\mu}G_{F}^{-1}(K-Q) \\
\label{eq:Ward_energy_Euclidean_2}
Q_{\nu}{\cal D}_{B\; \alpha\beta}^{\mu\nu}(K,K-Q) & = & (K^{\mu}-Q^{\mu})G_{B\;\alpha\beta}^{-1}(K) - K^{\mu}G_{B\; \alpha\beta}^{-1}(K-Q)
\end{eqnarray}
where ${\cal D}_{B/F}$ are the bosonic/fermionic amputated effective vertices defined in Sect.~\ref{sec:Transport_QFT}.  To go from Euclidean space to Minkowski space, we need to analytically continue $K$ and $Q$ towards real energies.  The proper choice here is dictated by the physics of transport coefficients encoded in the Kubo relation.  First, the Wightman correlator in Eq.~(\ref{eq:Kubo_relations}) can be expressed as the imaginary part of a retarded correlator.  Second, in the pinch limit, the two side rail propagators that connect the ladder kernel to the effective vertex have different boundary conditions, namely $G_{F}^{\rm ret}(k)G_{F}^{\rm adv}(k-q)$ or $G_{F}^{\rm adv}(k)G_{F}^{\rm ret}(k-q)$.  These two requirements uniquely fix the analytic continuation to $K \rightarrow k^{0} + i\epsilon$, $K-Q \rightarrow k^{0} - q^{0} - i\epsilon$ and $Q \rightarrow q^{0} + 2i\epsilon$.  Taking the pinch limit ($q^{0} \rightarrow 0$, ${\bf q} \rightarrow 0$) after doing the analytic continuation and using $G_{F\;\rm ret/adv}^{-1}(p) = [\gamma^{0}(p^{0}\pm i\Gamma_{p}/2) - {\bf\gamma} \cdot {\bf p}]$ and $G_{B\;\alpha\beta}^{-1\;\rm ret/adv}(p) = g_{\alpha\beta}[(p^{0}\pm i\gamma_{p}/2)^{2} - |{\bf p}|^{2}]$ (valid when $p$ is nearly on-shell), Eqs.~(\ref{eq:Ward_energy_Euclidean_1})-(\ref{eq:Ward_energy_Euclidean_2}) become
\begin{eqnarray}
\label{eq:Ward_identity_1}
\lim_{q \rightarrow 0}\; q_{\nu}{\cal D}_{F}^{\mu\nu}(k + i\epsilon,k-q - i\epsilon) & = & ik^{\mu}\gamma^{0}\Gamma_{k} \;=\; 2ik^{\mu}\Sigma_{I}^{\rm ret}({\bf k}) \\
\label{eq:Ward_identity_2}
\lim_{q \rightarrow 0}\; q_{\nu}{\cal D}_{B\; \alpha\beta}^{\mu\nu}(k + i\epsilon,k-q - i\epsilon) & = & 2ik^{\mu}g_{\alpha\beta}\gamma_{k} \;=\; 2ik^{\mu}\Pi_{I\; \alpha\beta}^{\rm ret}({\bf k})
\end{eqnarray}
The last equality is valid near $k^{0}\approx |\mathbf{k}|$.  These last equations relate the effective vertex of the integral equation to the imaginary part of the on-shell retarded self-energy in the limit relevant to transport in the case of an energy-momentum tensor insertion.  Note that we used fully resummed retarted/advanced propagators because of the need to regularize pinch singularities.


Let us now see what are the implications of the Ward-like identities~(\ref{eq:Ward_identity_1})-(\ref{eq:Ward_identity_2}) on the integral equations~(\ref{eq:integral_equation_schematic_1})-(\ref{eq:integral_equation_schematic_2}).  Starting from the Euclidean versions of Eqs.~(\ref{eq:integral_equation_schematic_1})-(\ref{eq:integral_equation_schematic_2}),
\begin{eqnarray}
\label{eq:integral_equation_Euclidean_1}
{\cal D}_{F}^{\mu\nu}(K,K-Q) & = & {\cal I}_{F}^{\mu\nu}(K,K-Q) + \int\frac{d^{4}P}{(2\pi)^{4}}\; {\cal K}_{(1)}(K,P,Q){\cal D}_{F}^{\mu\nu}(P,P-Q) \nonumber \\
                             &   & \hspace{1.1in} + \int\frac{d^{4}P}{(2\pi)^{4}}\; {\cal K}_{(2)}^{\alpha\beta}(K,P,Q){\cal D}_{B\;\alpha\beta}^{\mu\nu}(P,P-Q) \\
\label{eq:integral_equation_Euclidean_2}
{\cal D}_{B\;\alpha\beta}^{\mu\nu}(K,K-Q) & = & {\cal I}_{B\;\alpha\beta}^{\mu\nu}(K,K-Q) + \int\frac{d^{4}P}{(2\pi)^{4}}\; {\cal K}_{(3)\;\alpha\beta}(K,P,Q){\cal D}_{F}^{\mu\nu}(P,P-Q) \nonumber \\
                                          &   & \hspace{1.23in} + \int\frac{d^{4}P}{(2\pi)^{4}}\; {\cal K}_{(4)}(K,P,Q){\cal D}_{B\;\alpha\beta}^{\mu\nu}(P,P-Q)
\end{eqnarray}
we multiply both sides by $Q_{\nu}$ and do the analytic continuation that leads to pinch singularities.  Note that this step is delicate, since the sum over Matsubara frequencies must be done before the analytic continuation; the result is that the integral equation keeps its form in Minkowski space, as shown in \cite{Gagnon_Jeon_2007} for electrical conductivity.  After taking the $q\rightarrow 0$ limit, we use the Ward-like identities~(\ref{eq:Ward_identity_1})-(\ref{eq:Ward_identity_2}) on both sides of Eqs.~(\ref{eq:integral_equation_Euclidean_1})-(\ref{eq:integral_equation_Euclidean_2}).  We thus obtain:
\begin{eqnarray}
\label{eq:Integral_equation_constrained_1}
2k^{\mu}\Sigma_{I}^{\rm ret}({\bf k}) & = & \int\frac{d^{4}p}{(2\pi)^{4}}\; {\cal M}_{(1)}(k,p)\;p^{\mu}\rho_{F}^{+}(p) + \int\frac{d^{4}p}{(2\pi)^{4}}\; {\cal M}_{(2)}^{\alpha\beta}(k,p)\;p^{\mu}\rho_{B\;\alpha\beta}^{+}(p) \\
\label{eq:Integral_equation_constrained_2}
2k^{\mu}\Pi_{I\; \alpha\beta}^{\rm ret}({\bf k}) & = & \int\frac{d^{4}p}{(2\pi)^{4}}\; {\cal M}_{(3)\;\alpha\beta}(k,p)\;p^{\mu}\rho_{F}^{+}(p) + \int\frac{d^{4}p}{(2\pi)^{4}}\; {\cal M}_{(4)}(k,p)\;p^{\mu}\rho_{B\;\alpha\beta}^{+}(p)
\end{eqnarray}
where we have used the decomposition ${\cal K}(k,p,q) = {\cal M}(k,p,q)G(p)G(p-q)$ and $\rho_{B/F}^{+}(p) = i(G_{B/F}^{\rm ret}(p)-G_{B/F}^{\rm adv}(p))$.  Here $\Sigma_{I}^{\rm ret}({\bf k})$ and $\Pi_{I\; \alpha\beta}^{\rm ret}({\bf k})$ are the imaginary parts of full retarded self-energies and the ${\cal M}_{(i)}$'s represent coherent additions of rungs.  These last equations implement the constraints imposed by the Ward-like identities~(\ref{eq:Ward_identity_1})-(\ref{eq:Ward_identity_2}); since they are expressed only in terms of known quantities (self-energies and spectral densities), they also impose constraints on the ${\cal M}_{(i)}$'s.  Following the procedure in \cite{Gagnon_Jeon_2007}, it is possible to invert the system of equations~(\ref{eq:Integral_equation_constrained_1})-(\ref{eq:Integral_equation_constrained_2}) and express the ${\cal M}_{(i)}$'s in terms of functional derivatives of self-energies.  The result is (Lorentz indices for bosonic self-energies and spectral densities are not shown for simplicity):
\begin{eqnarray}
\label{eq:Constraint_M_1} q^{\mu}(1+b){\cal
M}_{(1)\;bf}(k,q) = k^{\mu}\left(\frac{\delta
(2\Sigma_{I\;bf}^{\rm ret}(k))}{\delta \rho_{F}^{+}(q)}\right) &
\rightarrow & q^{\mu}{\cal M}_{(1)}(k,q) \propto k^{\mu}\left(\frac{\delta
(2\Sigma_{I}^{\rm ret}(k))}{\delta \rho_{F}^{+}(q)}\right) \\
\label{eq:Constraint_M_2} q^{\mu}(1+f){\cal
M}_{(2)\;bf}(k,q) = k^{\mu}\left(\frac{\delta
(2\Sigma_{I\;bf}^{\rm ret}(k))}{\delta \rho_{B}^{+}(q)}\right) &
\rightarrow & q^{\mu}{\cal M}_{(2)}(k,q) \propto k^{\mu}\left(\frac{\delta
(2\Sigma_{I}^{\rm ret}(k))}{\delta \rho_{B}^{+}(q)}\right) \\
\label{eq:Constraint_M_3} q^{\mu}(1+b){\cal M}_{(3)\;bf}(k,q) = k^{\mu}\left(\frac{\delta
(2\Pi_{I\;bf}^{\rm ret}(k))}{\delta \rho_{F}^{+}(q)}\right) &
\rightarrow & q^{\mu}{\cal M}_{(3)}(k,q) \propto k^{\mu}\left(\frac{\delta
(2\Pi_{I}^{\rm ret}(k))}{\delta \rho_{F}^{+}(q)}\right) \\
\label{eq:Constraint_M_4} q^{\mu}(1+f){\cal
M}_{(4)\;bf}(k,q) = k^{\mu}\left(\frac{\delta (2\Pi_{I\;bf}^{\rm
ret}(k))}{\delta \rho_{B}^{+}(q)}\right) & \rightarrow & q^{\mu}{\cal M}_{(4)}(k,q)
\propto k^{\mu}\left(\frac{\delta (2\Pi_{I}^{\rm ret}(k))}{\delta
\rho_{B}^{+}(q)}\right) 
\end{eqnarray}
where the rung kernels and self-energies are expanded in terms of their number of bosonic ($b$) or fermionic ($f$) spectral densities: 
\begin{eqnarray}
{\cal M} = \sum_{b,f = 1}^{\infty}{\cal M}_{bf} & \;\;\;  & \Sigma_{I}^{\rm ret} = \sum_{b,f = 1}^{\infty}\Sigma_{I\;bf}^{\rm ret} \;\;\;\;\;\;\; \Pi_{I}^{\rm ret} = \sum_{b,f = 1}^{\infty}\Pi_{I\;bf}^{\rm ret}
\end{eqnarray}
These definitions mean that ${\cal M}$, $\Sigma_{I}^{\rm ret}$ and $\Pi_{I}^{\rm ret}$ are ``blobs'' containing diagrams of all orders; these diagrams can be reorganized in terms of their number of bosonic or fermionic spectral densities.  The proportionality relations are statements about the structure of the ${\cal M}_{(i)}$'s.
  
The diagrammatic interpretation of Eqs.~(\ref{eq:Constraint_M_1})-(\ref{eq:Constraint_M_4}) is quite natural.  Since the propagators~(\ref{eq:11_propagators})-(\ref{eq:21_propagators}) are all proportional to spectral densities, then functional derivatives with respect to $\rho_{B/F}$ can be interpreted as opening boson/fermion lines in a Feynman diagram.  Thus each rung kernel is obtained by opening bosonic/fermionic lines in the appropriate self-energy diagrams.  Equations~(\ref{eq:Constraint_M_1})-(\ref{eq:Constraint_M_4}) are the desired constraints on the rung kernels.

The above constraints implement the physics of transport coefficients, namely the appearance of pinch singularities due to the low frequency, low momentum limit.  They serve two main useful purposes.  First, they serve as a guide to find the necessary rungs in order to obtain the shear viscosity at the desired level of accuracy.  The recipe is simple: make a loop expansion of all the bosonic/fermionic self-energies of the theory, keep the self-energies up to the desired level of accuracy and then open them in all possible ways to get the rungs.  This recipe seems to work well in theories where only pinch singularities are present (such as scalar theories), although it is hard to say if it works at all orders, since there exists no transport coefficient calculations that go beyond leading order (except for large $N$ theories \cite{Aarts_Martinez_2004}).  When other singularities are present (such as collinear singularities), some rung resummation is necessary and a loop expansion is thus not sufficient; additional power counting arguments must be supplied in order to get leading order results.  We come back to these issues in more details in Sect.~\ref{sec:Power_counting}.


A constraints similar to Eqs.~(\ref{eq:Constraint_M_1})-(\ref{eq:Constraint_M_4}) has been obtained in the case of electric conductivity in QED \cite{Gagnon_Jeon_2007}.  In that context, the starting point to obtain the constraint is the usual Ward identity for a current insertion.  Since the Ward identity for a current insertion is a direct consequence of gauge symmetry, the resulting constraint is an implementation of gauge invariance for the electric conductivity calculation.  More precisely, the constraint tells us which self-energies must be resummed in the side rail propagators for each rung present in the integral equation kernel so as to keep everyting transverse.  In the shear viscosity case, the constraints~(\ref{eq:Constraint_M_1})-(\ref{eq:Constraint_M_4}) are obtained from spacetime symmetries, not gauge symmetries.  Thus the interpretation of these constraints as an implementation of gauge invariance is not direct.  We argue that the constraints~(\ref{eq:Constraint_M_1})-(\ref{eq:Constraint_M_4}) are sufficient to preserve gauge invariance.  A quick argument would be that the constraints~(\ref{eq:Constraint_M_1})-(\ref{eq:Constraint_M_4}) have the same form as the one for electrical conductivity \cite{Gagnon_Jeon_2007}, so they should play the same role.  We show in more details in Appendix~\ref{app:Gauge_invariance} that the shear viscosity is indeed gauge parameter independent.





We note that expressions similar to Eqs.~(\ref{eq:Constraint_M_1})-(\ref{eq:Constraint_M_4}) are obtained using 2PI effective actions methods \cite{Aarts_Martinez_2003,Aarts_Martinez_2005}, with notable differences.  In 2PI methods, the ``constraint'' is in coordinate space and comes naturally from standard functional relations.  The important point is that the kernel of the integral equation is given by the functional derivative of the self-energy with respect to a dressed 2-point function, where the self-energy is itself given by the functional derivative of all amputated 2PI diagrams with respect to dressed 2-point functions.  No reference is made to the low frequency, low momentum limit or pinch singularities.  We also mention that there seems to be gauge invariance issues with 2PI methods because of the need to truncate the 2PI effective action \cite{Arrizabalaga_Smit_2002,Mottola_2003,Carrington_etal_2005} (see also the developments in \cite{Berges_2004,Reinosa_Serreau_2006}).  In contrast, our method does not make any reference to 2PI effective actions and start directly from symmetry principles, the Ward identity being the expression of these symmetries for quantum mechanical amplitudes.  It is more specific in the sense that the physics of transport coefficients is necessary here to obtain our constraint.


\section{Power Counting}
\label{sec:Power_counting}

In this section, we present power counting arguments to supplement the constraints~(\ref{eq:Constraint_M_1})-(\ref{eq:Constraint_M_4}).  We separate the analysis in two parts.  In the first part, we follow the recipe used in \cite{Gagnon_Jeon_2007} and outlined in Sect.~\ref{sec:Ward_identity}, i.e. we expand the self-energies of the theory, keep only the leading order ones and open them according to Eqs.~(\ref{eq:Constraint_M_1})-(\ref{eq:Constraint_M_4}).  We then use power counting arguments to verify that the rungs obtained contribute at leading order.  In the second part, we show that the naive expansion in terms of coupling constants is not sufficient when collinear singularities are present.  As pointed out in Sect.~\ref{sec:Transport_QFT}, a restricted but infinite class of diagrams must be resummed in order to get leading order results.  We again use power counting arguments to identify this class of diagrams.  For the rest of this paper, we divide the rungs in two categories, the ones containing collinear singularities (${\cal N}$) and those that do not (${\cal M}$).

\subsection{Power counting without collinear singularity (${\cal M}$)}
\label{sec:Power_counting_without_collinear}

\begin{figure}
\resizebox{\textwidth}{!}{\includegraphics{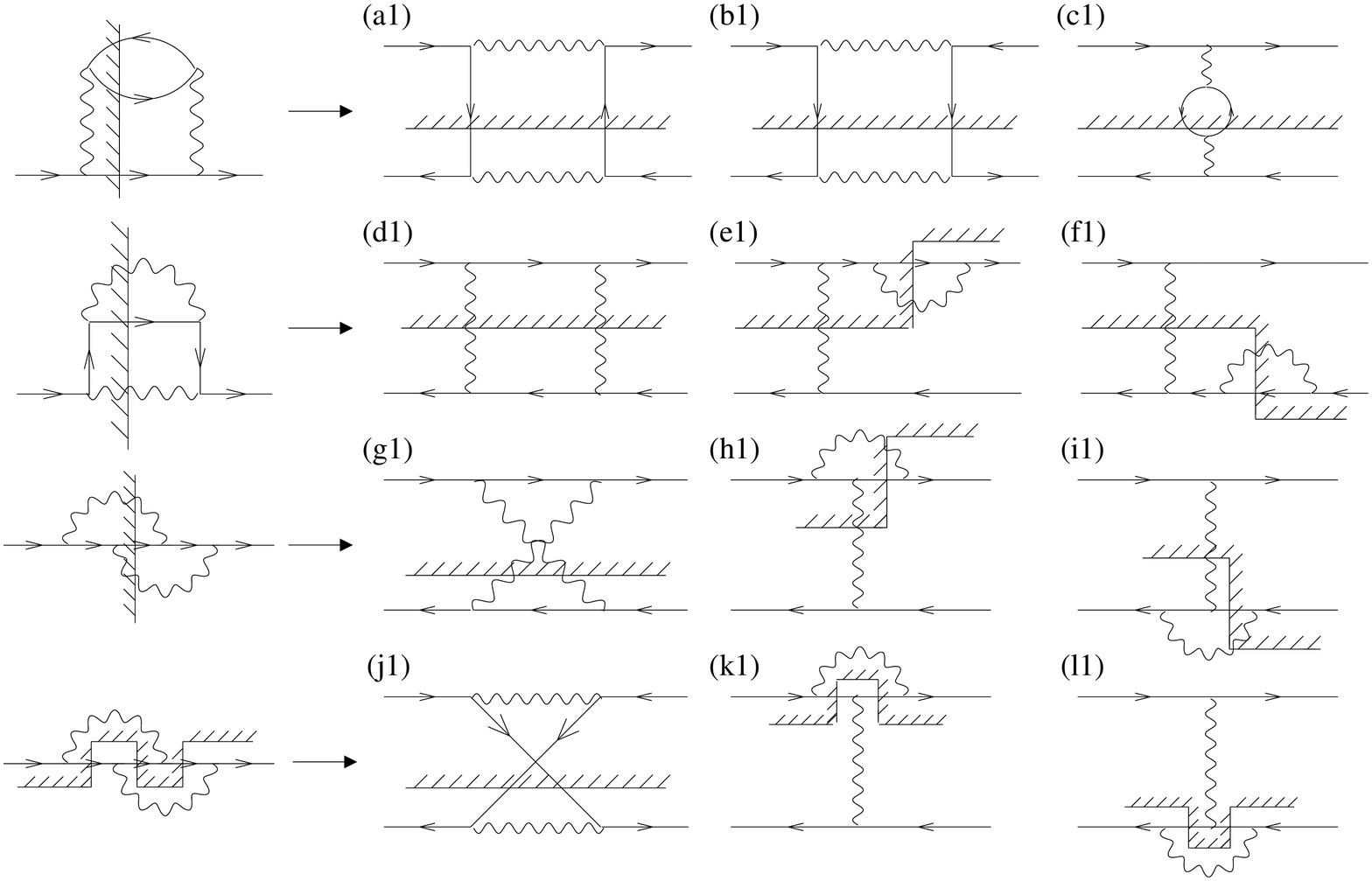}}
\caption{\label{fig:2to2_rungs_M1} Rungs of ${\cal M}_{(1)}$, obtained by functionally differentiating with respect to a fermionic spectral density the electron self-energies in the left column (c.f. Eq.~(\ref{eq:Constraint_M_1})).  Power counting arguments (see Ref.~\cite{Gagnon_Jeon_2007}) show that all these rungs contribute at leading order except diagrams (e1) and (f1), because massless three-body on-shell decays are suppressed.  Note that rungs (h1)-(i1) and (k1)-(l1) are different cuts of the same topology; any cut can be used for power counting purposes because, in the end, all possible cuts of a given topology must be included in the integral equation (c.f. Eqs.~(\ref{eq:integral_equation_1})-(\ref{eq:integral_equation_2})).  Since cut propagators represent on-shell excitations, all rungs (except rungs (e1),(f1),(h1),(i1)) are equivalent to $2\rightarrow 2$ scatterings with a soft exchange.  Note also that in a certain kinematical regime, rung topologies (g1)-(i1) contain collinear singularities and can be converted into $1\rightarrow 2$ collinear scatterings (we come back to this issue in Sects.~\ref{sec:Power_counting_with_collinear} and \ref{sec:Integral_equation_with_collinear}).  Figure taken from \cite{Gagnon_Jeon_2007}.}
\end{figure}

We first consider the case where the rungs do not contain collinear singularities.  As can be seen from Eqs.~(\ref{eq:integral_equation_schematic_1})-(\ref{eq:integral_equation_schematic_2}), there are four different types of rung kernels ${\cal M}_{(i)}$.  We treat them separately in the following.  According to Eq.~(\ref{eq:Constraint_M_1}), the rung kernel ${\cal M}_{(1)}$ is given by opening electron propagators in the imaginary part of the retarded self-energy of the electron.  At one-loop, the imaginary part of the electron self-energy is zero since an on-shell massless excitation cannot decay into two on-shell massless excitations; it is thus necessary to go to two loops for a leading order result (note that one-loop imaginary self-energies are possible when the photon is exactly collinear to the electron, a case we consider in Sect.~\ref{sec:Power_counting_with_collinear}).  The two-loop imaginary self-energies of the electron with their corresponding rungs are shown in Fig.~\ref{fig:2to2_rungs_M1}.  Note that there are many more cuts that correspond to imaginary two-loop self-energies.  We consider these other cuts when writing down the integral equation for the effective vertices (c.f. Eqs.~(\ref{eq:integral_equation_1})-(\ref{eq:integral_equation_2})); for the moment, we are only interested in the rung topology.  The power counting for the rungs in Fig.~\ref{fig:2to2_rungs_M1} is done in Ref.~\cite{Gagnon_Jeon_2007}.  The result is that the rungs corresponding to Coulomb scattering ((a1),(b1),(c1)) are $O(e^{2})$ and the others ((d1),(g1),(j1),(k1),(l1)) are $O(e^{4})$ when the off-shell exchange momentum is soft; but due to a partial cancellation between $2\rightarrow 2$ processes that do not change species types, all rungs are effectively $O(e^{4})$.

\begin{figure}
\resizebox{5in}{!}{\includegraphics{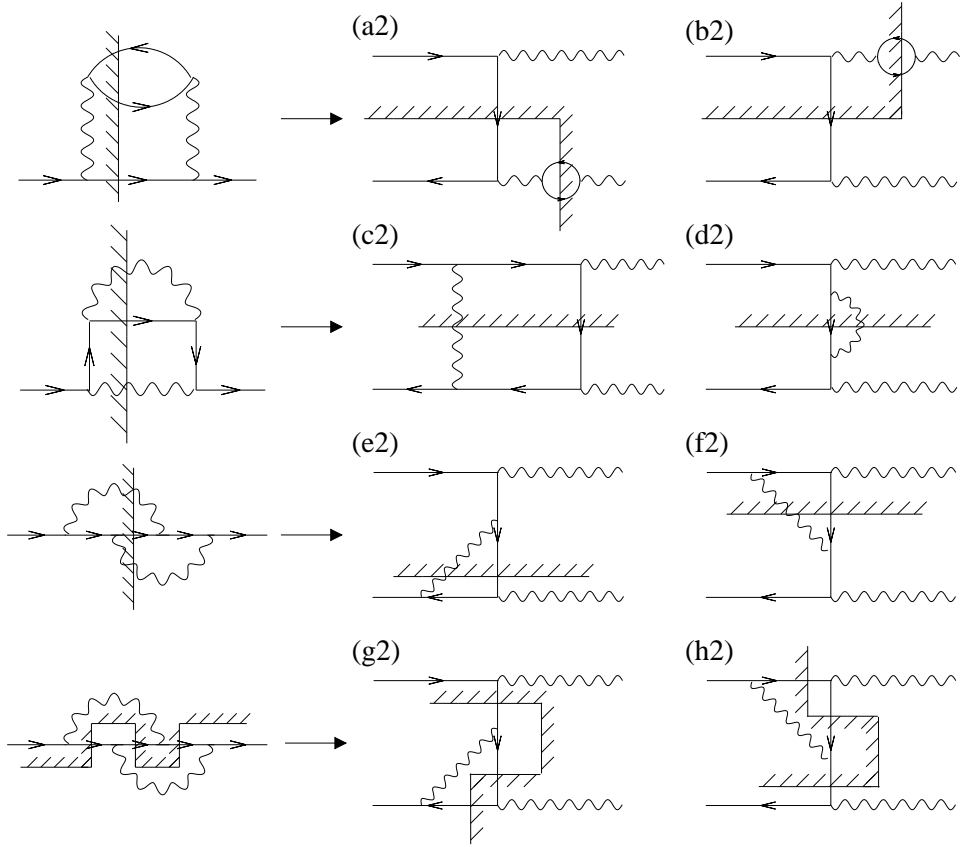}}
\caption{\label{fig:2to2_rungs_M2} Rungs of ${\cal M}_{(2)}$, obtained by functionally differentiating with respect to a bosonic spectral density the electron self-energies in the left column (c.f. Eq.~(\ref{eq:Constraint_M_2})).  Note that (e2)-(g2) and (f2)-(h2) are different cuts of the same topologies; any cut can be used for power counting purposes because, in the end, all possible cuts of a given topology must be included in the integral equation (c.f. Eqs.~(\ref{eq:integral_equation_1})-(\ref{eq:integral_equation_2})).  Except for rungs (a2) and (b2) containing suppressed three-body decays, they all contribute at leading order for shear viscosity and correspond to $2\rightarrow 2$ scattering processes.  Note that in a certain kinematical regime, rung topologies (e2)-(h2) contain collinear singularities and can be converted into $1\rightarrow 2$ collinear scattering processes (we come back to this issue in Sects.~\ref{sec:Power_counting_with_collinear} and \ref{sec:Integral_equation_with_collinear}).}
\end{figure}

\begin{figure}
\resizebox{5in}{!}{\includegraphics{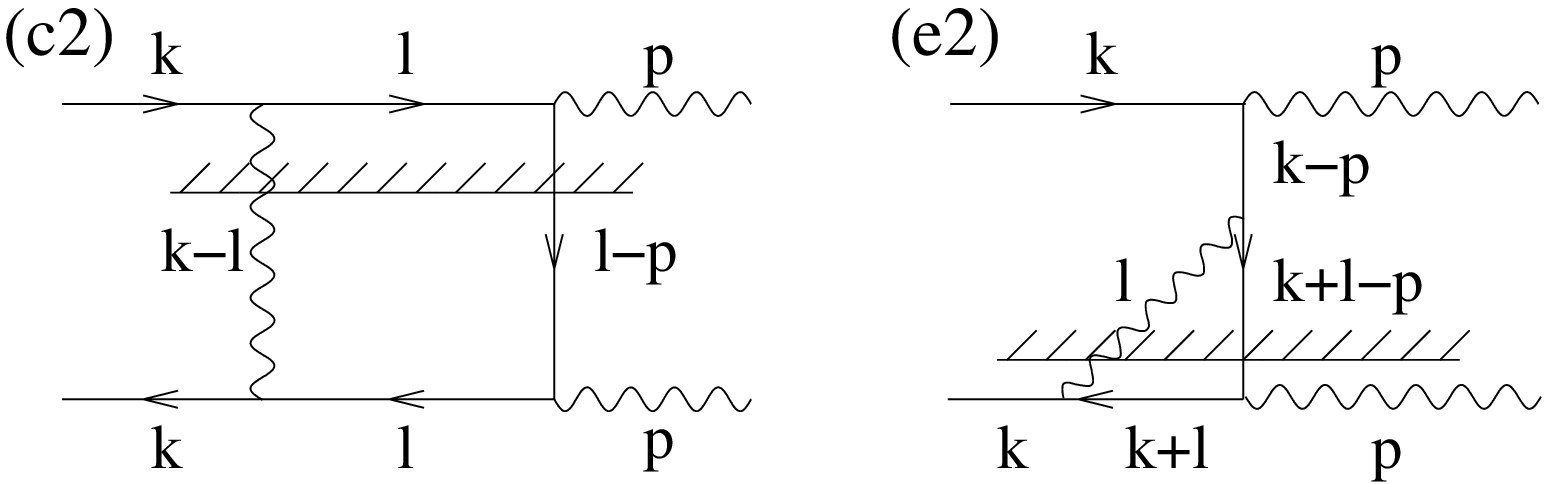}}
\caption{\label{fig:rungs_c2_e2} Momentum labels used to do the power counting of rungs (c2) and (e2) (c.f. Fig.~\ref{fig:2to2_rungs_M2}).}
\end{figure}

The rung kernel ${\cal M}_{(2)}$ is obtained by opening photon propagators in the same electron self-energies used for ${\cal M}_{(1)}$.  The resulting rungs are shown in Fig.~\ref{fig:2to2_rungs_M2}.  Let's estimate the size of rung (c2), reproduced with momentum labels in Fig.~\ref{fig:rungs_c2_e2}.  The expression for the rung is:
\begin{eqnarray}
\label{eq:Rung_c2}
{\cal M}_{(c2)} & = & -ie^{4}\int\frac{d^{4}l}{(2\pi)^{4}} \Delta_{B}^{-}(k-l)\Delta_{B}^{-}(l-p) G_{F}^{*}(l)G_{F}(l)
\end{eqnarray}
where for clarity we omitted Dirac matrices and Lorentz indices (irrelevant for noncollinear power counting).  If all momenta are hard, then (c2) is $O(e^{4})$ (this is true of all the rungs considered here).  The size of the rung is the same when the loop momentum is soft.  To see this, consider the momentum regime where $k$, $p$ are hard and on-shell while $l$ is soft and off-shell.  Using the delta functions inside the cut propagators to do two integrals over $l$, we get a $d^{2}l \sim O(e^{2}T^{2})$ suppression from phase space.  This phase space suppression is compensated by the two fermionic propagators $G_{F}(l)$, which are both $O(e^{-1}T^{-1})$ in size when $l$ is soft.  Rung (c2) is thus $O(e^{4})$ for both hard and soft loop momentum.  The size of rung (d2) is obtained similarly and is also $O(e^{4})$.

The power counting of rungs (e2) and (f2) is done in the same way.  For example, let's consider rung (e2).  The expression for the rung is (the momentum labels are shown in Fig.~\ref{fig:rungs_c2_e2}):
\begin{eqnarray}
\label{eq:Rung_e2}
\int\frac{d^{4}p}{(2\pi)^{4}}{\cal M}_{(e2)} = -ie^{4}\int\frac{d^{4}p}{(2\pi)^{4}}\int\frac{d^{4}l}{(2\pi)^{4}} \Delta_{B}^{+}(l)\Delta_{F}^{-}(k+l-l) G_{F}^{*}(k-p)G_{F}(k+l)
\end{eqnarray}
where we explicitly write the integration over $p$ coming from the integral equation (c.f. Eqs.~(\ref{eq:integral_equation_schematic_1})-(\ref{eq:integral_equation_schematic_2})).  In the momentum range where $k$, $p$, $l$ are hard and on-shell (this last condition being automatically enforced by the delta function in $\Delta_{B}^{+}(l)$), the fermionic propagators are generically $O(T^{-2})$.  With the additional requirements that $k\cdot p \sim k\cdot l \sim O(e^{2}T^{2})$, we have that $(k-p)^{2}\sim (k+l)^{2}\sim O(e^2 T^{2})$ and both propagators are $O(e^{-2}T^{-2})$.  On the other hand, these additional requirements are equivalent to $\theta_{kp}\sim \theta_{kl}\sim O(e)$ (where $\theta_{ab}$ is the angle between momenta $a$ and $b$), implying that phase space is restricted such that $d^{3}p \sim |{\bf p}|^{2}\sin \theta_{kp} d|{\bf p}|d\theta_{kp}d\phi \sim O(e^{2}T^{3})$ (and similarly for $d^{3}l$).  Combining all factors, we see that rung (e2) is $O(e^{4})$ in all momentum regimes where the exchange propagators are off-shell.  Note that this topology also contains collinear singularities in a certain kinematical regime; the consequences of this fact are explored in Sect.~\ref{sec:Power_counting_with_collinear}.  


\begin{figure}
\resizebox{\textwidth}{!}{\includegraphics{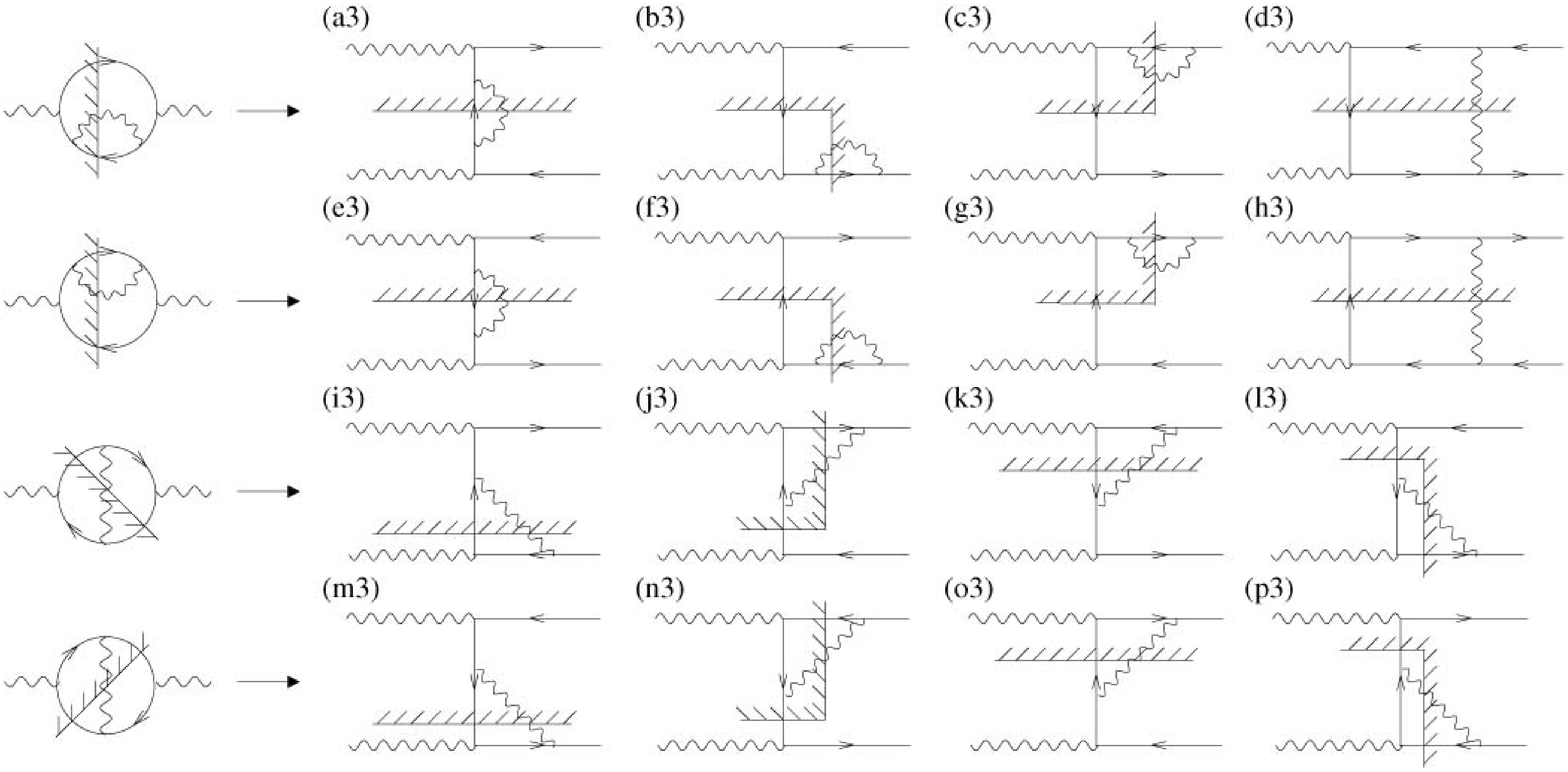}}
\caption{\label{fig:2to2_rungs_M3} Rungs of ${\cal M}_{(3)}$, obtained by functionally differentiating with respect to a fermionic spectral density the bosonic self-energies in the left column (c.f. Eq.~(\ref{eq:Constraint_M_3})).  Except for rungs (b3)-(c3) and (f3)-(g3) containing suppressed three-body decays, they all contribute at leading order for shear viscosity and correspond to $2\rightarrow 2$ scattering processes.  Note that in a certain kinematical regime, rung topologies (i3)-(p3) contain collinear singularities and can be converted into $1\rightarrow 2$ collinear scattering processes (we come back to this issue in Sects.~\ref{sec:Power_counting_with_collinear} and \ref{sec:Integral_equation_with_collinear}).}
\end{figure}

The analysis of the rung kernel ${\cal M}_{(3)}$ is similar to ${\cal M}_{(2)}$.  Figure~\ref{fig:2to2_rungs_M3} shows the rungs contained in ${\cal M}_{(3)}$ obtained by opening fermion propagators in the imaginary part of the retarded self-energy of the photon (see Eq.~(\ref{eq:Constraint_M_3})).  To do the power counting of ${\cal M}_{(3)}$, notice that the rungs in ${\cal M}_{(2)}$ and ${\cal M}_{(3)}$ are ``mirror images'' of each other, i.e. the incoming/outgoing electrons on the left of the rungs in ${\cal M}_{(2)}$ become outgoing/incoming electrons on the right in ${\cal M}_{(3)}$ (similarly for the photons).  Exception to this rule are rungs (a2)-(b2), (b3)-(c3), (f3)-(g3); they are however suppressed by three-body decays and are not included in the analysis.  Note also that ${\cal M}_{(3)}$ contains twice as many rungs compared to ${\cal M}_{(2)}$.  This is due to the fact that both electrons and positrons can flow in the RHS fermion loop in ${\cal M}_{(3)}$, whereas it is only a photon loop in ${\cal M}_{(2)}$.  Coming back to the power counting, we can see that all the power counting done for ${\cal M}_{(2)}$ can be applied to ${\cal M}_{(3)}$ since they both share the same ``mirror image'' rungs.

\begin{figure}
\resizebox{4in}{!}{\includegraphics{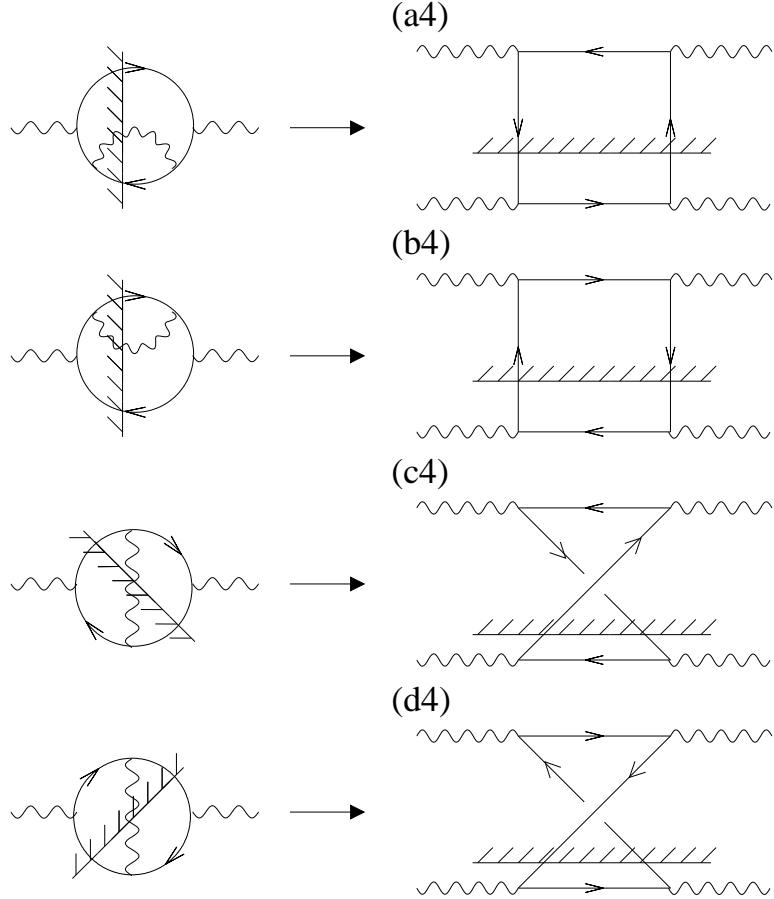}}
\caption{\label{fig:2to2_rungs_M4} Rungs of ${\cal M}_{(4)}$, obtained by functionally differentiating with respect to a bosonic spectral density the bosonic self-energies in the left column (c.f. Eq.~(\ref{eq:Constraint_M_4})).  All the rungs contribute at leading order for shear viscosity and correspond to $2\rightarrow 2$ scattering processes.  Note that none of these rungs enter the (leading order) collinear analysis of Sect.~\ref{sec:Power_counting_with_collinear}.}
\end{figure}

Figure~\ref{fig:2to2_rungs_M4} shows the rungs in ${\cal M}_{(4)}$, obtained by opening photon propagators in the imaginary part of the retarded self-energy of the photon (see Eq.~(\ref{eq:Constraint_M_4})).  The expressions for the rungs (a4)-(d4) are almost identical to the ones of (c2) and (e2); the power counting is thus very similar and gives $O(e^{4})$ for all four rungs.  This is to be expected, since there are no Coulomb divergence (i.e. no soft boson exchange) in (a4)-(d4) that would make the rungs $O(e^{2})$.  Note that none of the rungs in Fig.~\ref{fig:2to2_rungs_M4} are included in the (leading order) collinear analysis of Sect.~\ref{sec:Power_counting_with_collinear}, since they all pick up extra factors of coupling constants in the collinear regime.  We come back on this issue in the next section.

%

\subsection{Power counting with collinear singularities (${\cal N}$)}
\label{sec:Power_counting_with_collinear}

The power counting with collinear singularities is more subtle than the one done in the previous section, due to the additional restrictions to a particular kinematical regime.  The following discussion is largely based on the work of Arnold, Moore and Yaffe (in particular Ref.~\cite{AMY_2001a}).

In Ref.~\cite{Gagnon_Jeon_2007}, we use the ``1-2'' formalism to estimate the size of all rungs.  However, the 1-2 formalism is not the most appropriate tool for power counting, especially when collinear singularities are involved.  For instance, all four ``1-2'' propagators have a temperature dependent part that can dominate in certain kinematical regimes.  In comparison, the Keldysh basis (c.f. Eq.~\ref{eq:physical_functions}) is cleaner, since only the $G_{rr}$ contains distribution functions; it is thus easier to spot Bose-Einstein enhancements when momenta are soft and evaluate the size of each propagator.  The Keldysh formalism has another advantage compared to the ``1-2'' formalism.  The fact that the Keldysh basis is simpler (i.e. absence of $aa$ propagators, absence of vertices with an odd number of $a$'s) combined with the easiness of identifying pinching contributions (i.e. $G_{ra}G_{ar}$ or $G_{ar}G_{ra}$) makes it very convenient.  In particular, it renders apparent some near cancellations that are otherwise very hard to see in the ``1-2'' basis.  We thus use the Keldysh basis to identify leading order rung topologies in the following; once one is found (and there are no near cancellations), we can include it (and all its cuts) in the final integral equation.

\begin{figure}
\resizebox{\textwidth}{!}{\includegraphics{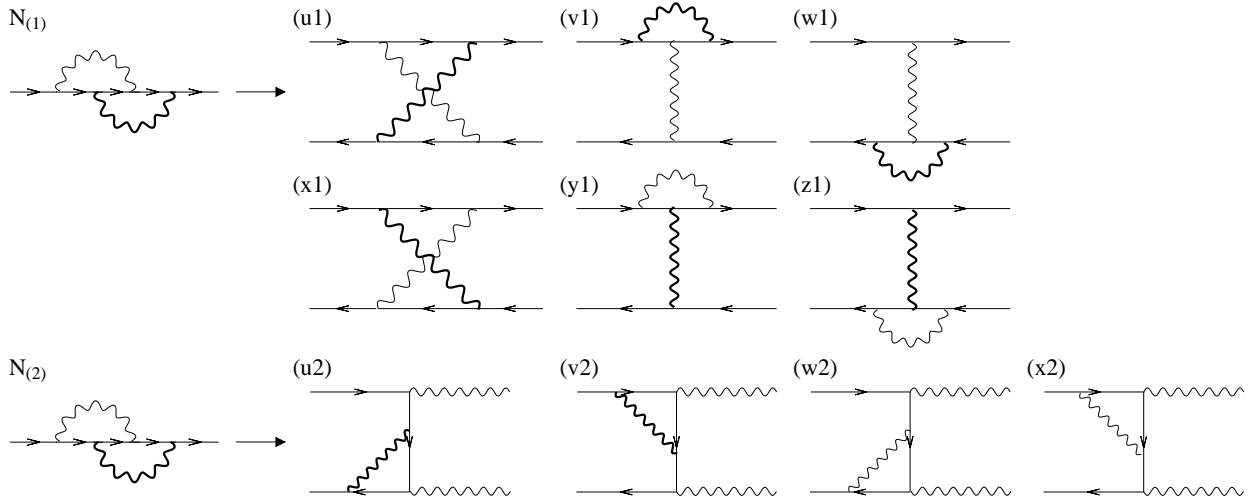}}
\caption{\label{fig:1to2_rungs_fermionic} Rungs of ${\cal N}_{1}$ and ${\cal N}_{2}$ obtained from the fermionic self-energies in the left column by functional differentiation with respect to $\rho_{F}$ and $\rho_{B}$, respectively.  A thick line corresponds to a soft propagator.  From power counting arguments, it can be shown that only rungs (u1), (x1), (u2), (v2) contribute at leading order and must be included in the collinear analysis.}
\end{figure}

\begin{figure}
\resizebox{\textwidth}{!}{\includegraphics{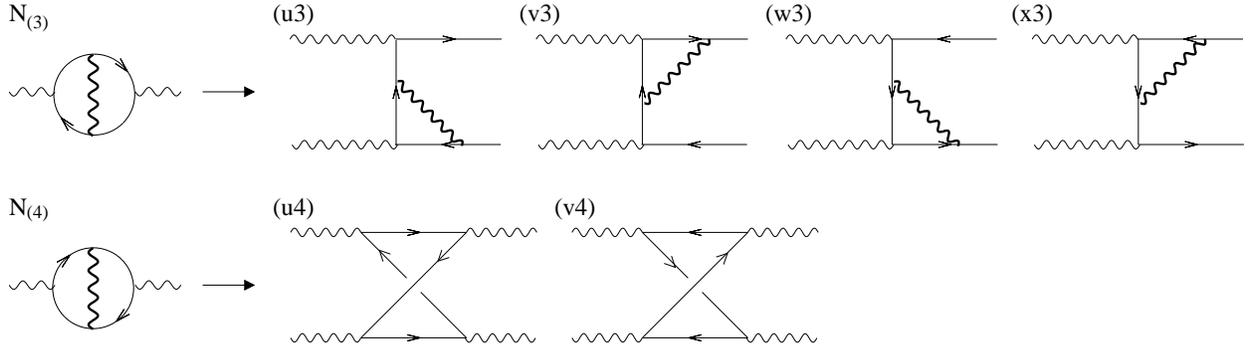}}
\caption{\label{fig:1to2_rungs_bosonic} Rungs of ${\cal N}_{3}$ and ${\cal N}_{4}$ obtained from the bosonic self-energies in the left column by functional differentiation with respect to $\rho_{F}$ and $\rho_{B}$, respectively.  A thick line corresponds to a soft propagator.  From power counting arguments, it can be shown that only rungs (u3)-(x3) contribute at leading order and must be included in the collinear analysis.}
\end{figure}

The first step is to identify the leading order fermionic and bosonic self-energies when the initial hard electron or photon is in the collinear regime.  As explained in Sect.~\ref{sec:Transport_QFT}, collinear singularities occur when a retarded and an advanced propagator with nearly collinear momenta are multiplied together.  For power counting purposes, we also need at least one of the photon propagator to be $rr$ (we come back to this point below).  By inspection, we see that only the self-energies shown in Figs.~\ref{fig:1to2_rungs_fermionic}-\ref{fig:1to2_rungs_bosonic} satisfy these criteria (note that the photon self-energies in Fig~\ref{fig:1to2_rungs_bosonic} are identical to the ones of \cite{AMY_2001a}).  The rest of this section is devoted to the power counting of the resulting rungs.


\begin{figure}
\resizebox{3in}{!}{\includegraphics{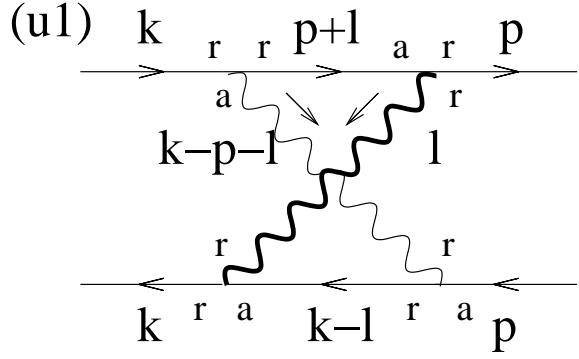}}
\caption{\label{fig:Rung_u1} Momentum and $r$,$a$ labels used to do the power counting of rungs (u1) (c.f. Fig.~\ref{fig:1to2_rungs_fermionic} and Eq.~(\ref{eq:Rung_u1})).  A thick line corresponds to a soft propagator.}
\end{figure}

The first class of rungs comes from opening fermionic self-energies (see Fig.~\ref{fig:1to2_rungs_fermionic}).  Rung (u1) is reproduced with $r$,$a$ indices and momentum labels in Fig.~\ref{fig:Rung_u1} (rung (x1) is done in a similar way).  The expression corresponding to rung (u1) is:
\begin{eqnarray}
\label{eq:Rung_u1}
\int\frac{d^{4}p}{(2\pi)^{4}}{\cal N}_{(u1)}{\cal D}_{F}(p) & \sim & e^{4}\int\frac{d^{4}p}{(2\pi)^{4}}\int\frac{d^{4}l}{(2\pi)^{4}} G_{B\;\alpha\beta}^{ar}(k-p-l)G_{B\;\mu\nu}^{rr}(l) \nonumber \\
                                                            &   & \hspace{0.5in} \times \gamma^{\alpha}G_{F}^{ra}(p+l)\gamma^{\mu} {\cal D}_{F}(p) \gamma^{\beta}G_{F}^{ra}(k-l)\gamma^{\nu}
\end{eqnarray}
where we explicitly write the integral over $p$ and the effective vertex coming from the integral equation (c.f. Eq.~\ref{eq:integral_equation_schematic_1}).  Note that there are other possible $r$,$a$ assigments (this is why we put a proportionality sign in Eq.~(\ref{eq:Rung_u1})): some of them are subleading, some of them contribute at leading order, but the important point is that no near cancellations between different $r$,$a$ configurations are at work for this rung.  The external legs are hard and nearly on-shell due to pinch singularities, i.e. $k\sim p\sim T$ and $k^{2}\sim p^{2}\sim O(e^{2}T^{2})$.  We consider the case where the loop momentum $l$ is soft (the case where $l$ is hard corresponds to the non-collinear rung (g1)).  The integral over $dl^{0}$ is dominated by the kinematical range where both fermionic propagators are collinearly singular; this happens over the parametrically small frequency width $dl^{0}\sim O(e^{2}T)$ \cite{AMY_2001a} and when $(p+l)^{2}\sim (k-l)^{2}\sim O(e^{2}T^{2})$ (or, equivalently, $\theta_{pl}\sim \theta_{kl}\sim O(e)$).  These last requirements are equivalent to the statement that the side rail electrons must be collinear with the hard photon.  In particular, $\theta_{pl}\sim O(e)$ implies that the phase space of $p$ is restricted to an $O(e^{2})$ region since $d^{3}p \sim |{\bf p}|^{2}\sin \theta_{pl} d|{\bf p}|d\theta_{pl}d\phi \sim O(e^{2}T^{3})$.

When $l$ is soft, $G_{B}^{rr}(l) = (1+2n_{B}(l^{0}))\rho_{B}(l)$ is HTL resummed and its size depends on the momentum flowing through it.  The dominant contribution comes from soft spacelike momenta.  In this situation, Landau damping gives rise to an $O(e^{2})$ imaginary self-energy; the size of the spectral density is thus $\rho_{B\; \rm HTL}(l) \sim \Pi_{I}(l)/(l^{2}+\Pi_{I}(l))^{2}\sim O(e^{-2})$ and the size of the propagator is $G_{B\; \rm HTL}^{rr}(l) \sim O(e^{-3})$.  Collecting all powers of $e$, we get $e^{4}\times e^{2}\times (e^{2}\times e^{3})\times e^{-3}\times (e^{-2}\times e^{-2})$, where the $e^{4}$ comes from the four explicit vertices, $e^{2}$ from the (small angle) restriction on the phase space of $p$, $(e^{2}\times e^{3})$ from the (parametrically small) integration over $l^{0}$ and the soft integration over ${\bf l}$, $e^{-3}$ from the soft HTL resummed $rr$ propagator and $(e^{-2}\times e^{-2})$ from the two collinearly singular fermionic propagators.  Rung (u1) is thus $O(e^{4})$ and contribute at leading order in this particular regime.

Rungs (y1)-(z1) and (v1)-(w1) have the same topology but are evaluated in different momentum regimes.  We argue that the momentum regime of rungs (y1)-(w1) (soft momentum exchange between the side rails) is suppressed compared to the case when the side rail momentum exchange is hard.  A quick argument to see this suppression is to notice that for any rung with incoming momentum $k$ and a soft exchange ($p\sim O(eT$)) between the side rails, we can expand the two pinching side rail propagators that connect the rung to the effective vertex in a Taylor series, i.e. $G(k+p)\approx G(k) + p^{\mu}\partial_{\mu}G(k)$.  The first term in the Taylor expansion gives rise to a rung with incoming and outgoing momenta $k$ and so does not disturb the ladder at all.  We are thus left with only the term proportional to $p$ that is $O(e)$ suppressed.  In other words, the change induced by adding a rung with a soft exchange between the side rails is parametrically smaller than the one caused by adding a rung with a hard exchange.  This argument is valid for all rungs with no hard exchange between the two side rails.  A better way to see this suppression is to look directly at the resulting linearized Boltzmann equation and see the cancellation between scattering processes for which both incident excitations undergo a soft scattering without changing species type \cite{AMY_2000,Gagnon_Jeon_2007}.

\begin{figure}
\resizebox{5in}{!}{\includegraphics{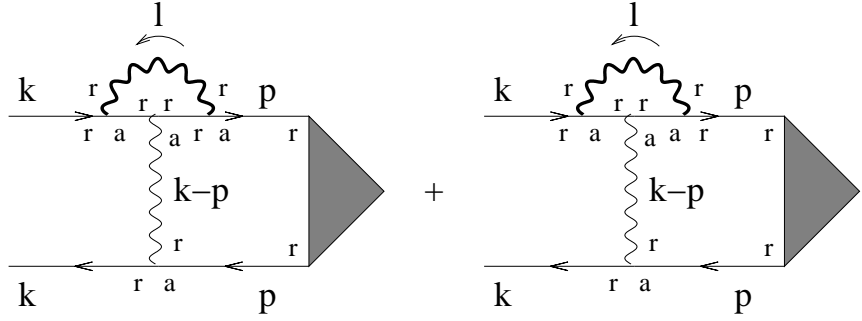}}
\caption{\label{fig:Cancellation_rung_v1} Possible assignments of $r$,$a$ labels for rung (v1); other assignments are subleading because of, e.g. the absence of pinching propagators inside the soft vertex correction.  Expanding the $rr$ propagators as $G_{B/F}^{rr}(k) = (1 \pm 2n_{B/F}(k^{0}))\left(G_{B/F}^{ra}(k)-G_{B/F}^{ar}(k)\right)$, we can show that these two contributions nearly cancel (c.f. Eq.~(\ref{eq:Cancellation_v1_1})).  Figure taken from \cite{Gagnon_Jeon_2007_erratum}.}
\end{figure}

The complete power counting of rungs (v1) and (w1) is done in Ref.\cite{Gagnon_Jeon_2007}.  The result is they are both $O(e^{4})$.  They should nevertheless not be included in a leading order analysis due to a near cancellation  \cite{Gagnon_Jeon_2007_erratum} \footnote{We thank S. Caron-Huot for pointing this cancellation to us.}.  The cancellation can be seen as follows.  The assigment of $r$,$a$ labels in rung (v1) is tightly constrained because of the need for the side rail propagators (connecting the vertex to the rung) to pinch, the need for the two fermion propagators inside the vertex correction to pinch and also the need for the soft bosonic vertex correction to be $rr$.  The two possible $r$,$a$ configurations of the ``rung-side rails-vertex'' system are shown in Fig.~\ref{fig:Cancellation_rung_v1}.  Adding the two contributions, we get:
\begin{eqnarray}
\label{eq:Cancellation_v1_1}
{\cal N}_{(v1)} & \sim & e^{4}\left(\mbox{Common part}\right)\left[G_{F}^{rr}(p+l)G_{F}^{ar}(p) + G_{F}^{ra}(p+l)G_{F}^{rr}(p)\right] \nonumber \\
              & \sim & e^{4}\left(\mbox{Common part}\right)\left[(1-2n_{F}(p^{0}+l^{0}))G_{F}^{ra}(p+l)G_{F}^{ar}(p) \right. \nonumber \\
              &      & \left. \hspace{1.2in} - (1-2n_{F}(p^{0}))G_{F}^{ra}(p+l)G_{F}^{ar}(p)\right]
\end{eqnarray}
where, to obtain the second line, we used $G_{F}^{rr}(p) = (1-2n_{F}(p^{0}))\left(G_{F}^{ra}(p)-G_{F}^{ar}(p)\right)$ and kept only the pinching part.  When $l$ is soft, the two contributions nearly cancel (up to $O(e)$ corrections) and make the rung subleading.

\begin{figure}
\resizebox{2.5in}{!}{\includegraphics{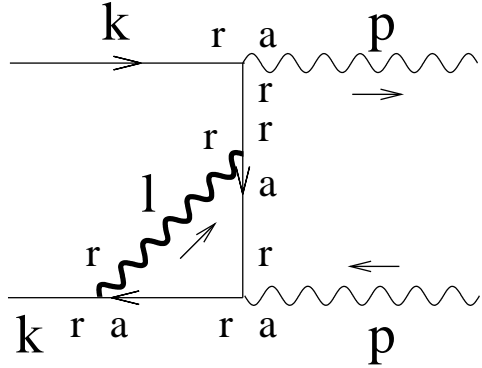}}
\caption{\label{fig:Rung_u2} Momentum and $r$,$a$ labels used to do the power counting of rungs (u2) (c.f. Fig.~\ref{fig:1to2_rungs_fermionic} and Eq.~(\ref{eq:Rung_u2})).  A thick line corresponds to a soft propagator.}
\end{figure}

Rung (u2) is reproduced with momentum and $r$,$a$ labels in Fig.~\ref{fig:Rung_u2}.  The corresponding expression is:
\begin{equation}
\label{eq:Rung_u2}
\int\frac{d^{4}p}{(2\pi)^{4}}{\cal N}_{(u2)} \sim e^{4}\int\frac{d^{4}p}{(2\pi)^{4}}\int\frac{d^{4}l}{(2\pi)^{4}} G_{B\;\mu\nu}^{rr}(l) \gamma^{\alpha}G_{F}^{rr}(k-p)\gamma^{\mu}G_{F}^{ar}(k-p+l)\gamma^{\beta}G_{F}^{ra}(k+l)\gamma^{\nu}
\end{equation}
where we explicitly write the integral over $p$ coming from the integral equation (c.f. Eq.~\ref{eq:integral_equation_schematic_1}).  The external legs are hard and nearly on-shell due to pinch singularities, i.e. $k\sim p\sim T$ and $k^{2}\sim p^{2}\sim O(e^{2}T^{2})$.  We consider the case where the loop momentum $l$ is soft (the case where $l$ is hard corresponds to the non-collinear rung (e2)).  Similarly to rung (u1), the integral over $dl^{0}$ is dominated by the region when the two propagators $G_{F}^{ar}(k-p+l)G_{F}^{ra}(k+l)$ nearly pinch: this happens when the hard photon is collinear with the hard electrons (i.e. when $\theta_{kp} \sim O(g)$).  To get a leading order contribution, we also need $l$ to be spacelike, giving $G_{B\; \rm HTL}^{rr}(l) \sim O(e^{-3})$.  Note that $G_{F}^{rr}(k-p)$ does not have any pole in $l^{0}$ and thus do not give any singular contribution to the integral.  Collecting all powers of $e$, we get $e^{4}\times e^{2}\times (e^{2}\times e^{3})\times e^{-3}\times (e^{-2}\times e^{-2}) \sim O(e^{4})$, as for rung (u1).  Thus rungs (u2)-(v2) must be included in the leading order collinear analysis.  Note that rungs (w2)-(x2) are already included in Sect.~\ref{sec:Power_counting_without_collinear} (compare with rungs (e2) and (f2)).


%

\begin{figure}
\resizebox{\textwidth}{!}{\includegraphics{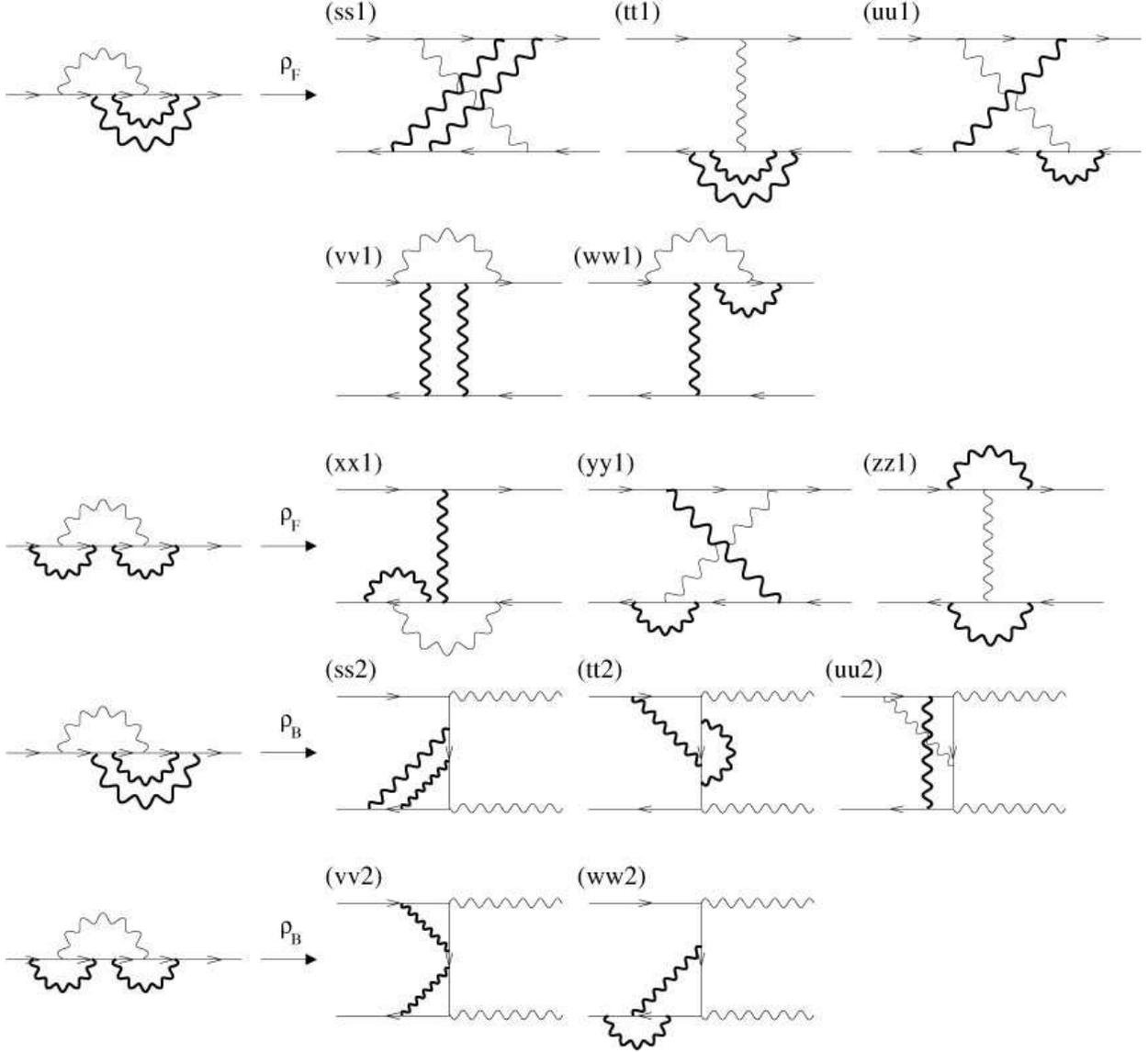}}
\caption{\label{fig:Rungs_LPM_fermionic} Possible 3-loop fermionic self-energies that contribute at leading order.  A thick line corresponds to a soft propagator.  The corresponding rungs (obtained by functionally differentiating the self-energies with respect to $\rho_{F}$ or $\rho_{B}$) are also shown.  Note that not all the possible rungs are shown, this is only a representative sample.  Power counting arguments tell us that only rungs (ss1) and (ss2) must be included in the collinear analysis.}
\end{figure}

More generally, power counting allows all fermionic self-energies with any number of vertex corrections to the hard photon to be included (some examples of self-energies are shown in Fig.~\ref{fig:Rungs_LPM_fermionic}, along with the corresponding rungs).  To see that, note that a vertex correction (on any end of the hard photon) adds two explicit vertices ($e^{2}$), an integral over soft momenta ($e^{3}$), two collinearly singular propagators ($e^{-2}$) and one soft $rr$ propagator ($e^{-3}$).  Thus adding a vertex correction amounts to an $O(e^{0})$ change to the self-energy.  This is true for any number of vertex corrections on both ends of the hard photon.

However, most of the rungs in Fig.~\ref{fig:Rungs_LPM_fermionic} do not contribute at leading order and can be left out of the analysis.  For instance, using arguments similar to the ones used for rungs (v1)-(w1), it is easy to show that cancellations between different $r$,$a$ assigments occur for rungs (tt1), (uu1), (yy1), (zz1), (vv2) and make them subleading.  Rungs (vv1), (ww1), (xx1) are also subleading since there is no hard exchange between the side rails (same argument as for rungs (y1)-(w1)).  Lastly, rungs (tt2), (uu2), (ww2) should not be considered in the first place, since they are obtained by opening fermionic self-energies with respect to soft HTL resummed spectral densities; this not allowed, since the functional derivative in Eq.~(\ref{eq:Constraint_M_2}) is with respect to $\rho_{B}$, not $\rho_{B\;\rm HTL}$.  Note that the above arguments are robust with respect to the addition of more vertex corrections to the self-energies.  Thus after a careful analysis, we see that all the rungs in Fig.~\ref{fig:Rungs_LPM_fermionic} are subleading, except rungs similar to rungs (u1), (x1), (u2) and (v2) with an arbitrary number of crossed soft photons.  This completes the power counting for the class of rungs obtained from fermionic self-energies.


%
%
%
%

\begin{figure}
\resizebox{\textwidth}{!}{\includegraphics{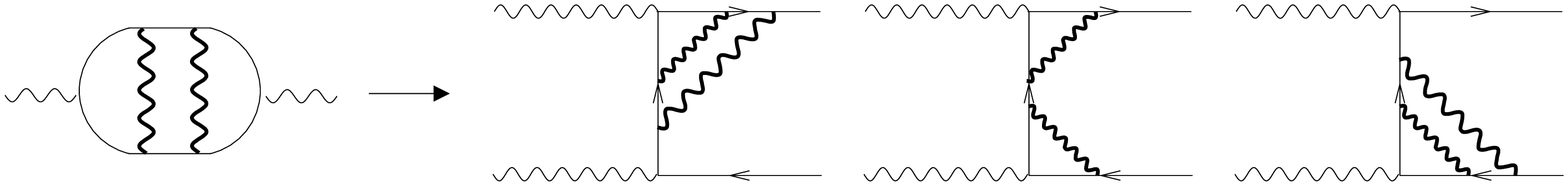}}
\caption{\label{fig:Rungs_LPM_bosonic} Possible 3-loop bosonic self-energy that contribute at leading order.  A thick line corresponds to a soft propagator.  The corresponding rungs (obtained by functionally differentiating the self-energy with respect to $\rho_{F}$) are also shown.  Note that not all the possible rungs are shown, this is only a representative sample.  Power counting arguments tell us that only the rungs with soft corrections to one vertex at a time must be included in the collinear analysis.}
\end{figure}

The second class of rungs comes from opening bosonic self-energies (see Fig.~\ref{fig:1to2_rungs_bosonic}).  We first notice that, as in the non-collinear case, rungs (u3)-(x3) are the ``mirror images'' of rungs (u2)-(v2).  Consequently, the power counting of rungs (u3)-(x3) is similar to rungs (u2)-(v2) and shows that they are all $O(e^{4})$.  As for fermionic self-energies, an infinite number of bosonic self-energies must also be included at leading order.  These self-energies are shown in Fig.~\ref{fig:Rungs_LPM_bosonic} and are analyzed in details in \cite{AMY_2001a}.  The key point is that the addition of soft exchange photons to the bosonic self-energies in Fig.~\ref{fig:Rungs_LPM_bosonic} are in fact corrections to the incoming/ougoing hard photon vertex; they are similar to the multiple vertex corrections in the fermionic self-energies (c.f. Fig.~\ref{fig:Rungs_LPM_fermionic}) and can be treated in the same way.

Lastly, rungs (u4), (v4) (c.f. Fig.~\ref{fig:1to2_rungs_bosonic}) are subleading and are not considered in the collinear analysis.  The power counting for these rungs is very similar to the one for (u1), (x1), with the difference that the two bosonic exchange propagators are replaced by two fermionic propagators.  From a power counting point of view, it implies that rungs (u4), (v4) do not have the Bose-Einstein enhancement associated with a soft bosonic exchange propagator and are thus subleading.  In any case, rungs (u4), (v4) are not allowed since they are obtained by opening bosonic self-energies with respect to $\rho_{B\;\rm HTL}$ and not $\rho_{B}$ as prescribed by Eq.~\ref{eq:Constraint_M_4}.

The infinite number of rung types with arbitrary number of soft vertex corrections must be resummed using another integral equation.  This is the origin of the comment made in Sect.~\ref{sec:Transport_QFT} about the necessity of using two integral equations embedded in each other to obtain a leading order result.  The fact that an arbitrary number of soft vertex corrections are allowed is a direct consequence of collinear singularities and is the diagrammatic implementation of the LPM effect \cite{AMY_2001a,Aurenche_etal_2000,Baym_etal_2006}.  We will see more clearly in Sect.~\ref{sec:Integral_equation_with_collinear} how these collinear rungs are related to the $1+N \rightarrow 2+N$ collinear scattering processes found in e.g. Ref.~\cite{AMY_2002}.







\section{Derivation of the Integral Equations}
\label{sec:Integral_equations}

\begin{figure}
\resizebox{5.05in}{!}{\includegraphics{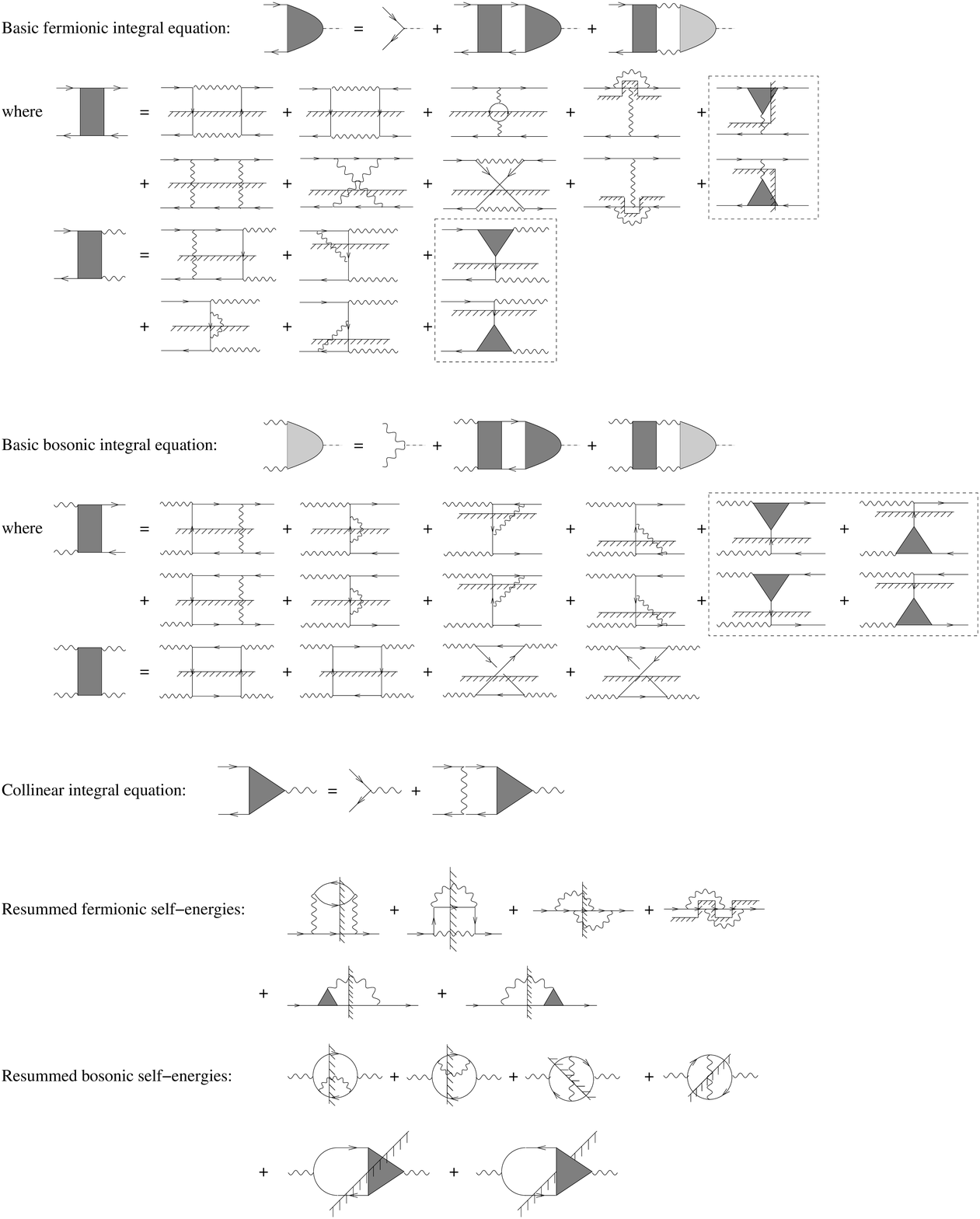}}
\caption{\label{fig:Summary_viscosity} Diagrammatic summary of our leading order calculation of shear viscosity in hot QED.  The basic integral equations (c.f. Eqs.~(\ref{eq:integral_equation_schematic_1})-(\ref{eq:integral_equation_schematic_2})) are due to the usual pinch singularities; their solutions (represented by a grey half circles) must be substituted in the initial Kubo relation to get the shear viscosity.  All the rungs included in the kernels of the basic integral equations (represented by grey rectangles) are shown.  These rungs can be divided in two categories: those corresponding to $2\rightarrow 2$ scatterings (located outside the dotted rectangles) and those corresponding to $1\rightarrow 2$ collinear scatterings (located inside the dotted rectangles).  The collinear rungs represent an infinite number of vertex corrections (represented by a grey triangle) that are resummed using the collinear integral equation (c.f. Eq.~(\ref{eq:Vertex_resummation})).  Also shown are the self-energies that must be resummed in the side rail propagators in order to satisfy the constraints due to the Ward-like identities~(\ref{eq:Constraint_M_1})-(\ref{eq:Constraint_M_4}).}
\end{figure}

A visual summary of the resummation program for the computation of shear viscosity at leading order, including all the necessary collinear and non-collinear rungs, is shown in Fig.~\ref{fig:Summary_viscosity}.  The next goal is to write down the appropriate integral equations (including all the cuts of the topologies shown in Fig.~\ref{fig:Summary_viscosity}) and show their equivalence to the results of Arnold, Moore and Yaffe \cite{AMY_2003a,AMY_2003b} obtained using effective kinetic theory.

The essential steps leading to the linearized Boltzmann equation for shear viscosity in scalar theories can be found in \cite{Jeon_1995,Jeon_Yaffe_1996}.  The same method (generalized to fermions and gauge theories) is used in \cite{Gagnon_Jeon_2007} to obtain the Boltzmann equation for electrical conductivity in hot QED.  Since the details of the method for both bosons and fermions are available in the litterature, we only outline the important steps in the present section.

As is done in Sect.~\ref{sec:Power_counting}, we separate the analysis of collinear and non-collinear rungs in Eqs.~(\ref{eq:integral_equation_schematic_1})-(\ref{eq:integral_equation_schematic_2}) according to ${\cal K} = ({\cal M}+{\cal N}){\cal F}$, where ${\cal N}$ and ${\cal M}$ correspond to the collinear and non-collinear rungs, respectively.  

%

\subsection{Integral equation without collinear singularity (${\cal M}$)}
\label{sec:Integral_equation_without_collinear}

\begin{figure}
\resizebox{\textwidth}{!}{\includegraphics{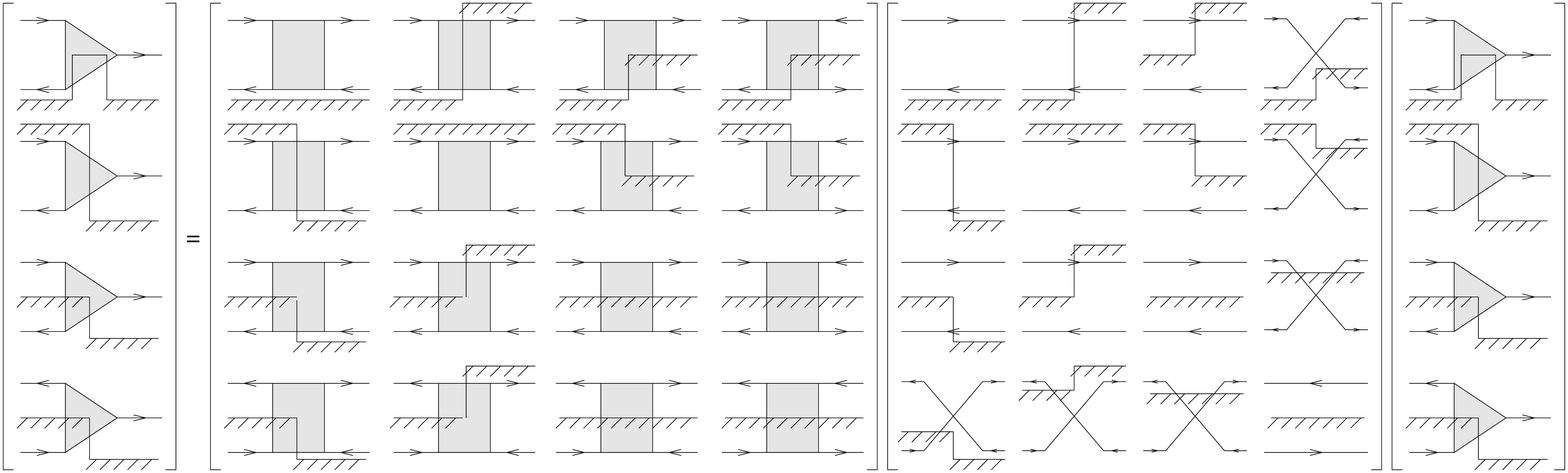}}
\caption{\label{fig:4x4_integral_equation} Graphical representation of the integral equations~(\ref{eq:integral_equation_1})-(\ref{eq:integral_equation_2}).  Only the term ${\cal K}_{1}$ is represented here; the three other terms have the same cut structure.  The inhomogeneous term $\mbox{\boldmath ${\cal I}$}(p)$ is not shown explicitly.  It is a column vector with only the first component being nonzero, since it is an operator insertion and is thus point-like ({\it i.e.} it can only be in the shade or out of the shade).  Figure taken from Ref.~\cite{Gagnon_Jeon_2007}.}
\end{figure}

We first consider the case with only non-collinear rungs for simplicity.  We want to write down Eqs.~(\ref{eq:integral_equation_schematic_1})-(\ref{eq:integral_equation_schematic_2}) in the ``1-2'' formalism, including all possible cuts of the effective vertices, the side rails and the rungs.  A compact way of doing that, pioneered by Jeon~\cite{Jeon_1995}, is to write the integral equations~(\ref{eq:integral_equation_schematic_1})-(\ref{eq:integral_equation_schematic_2}) in matrix form, where each component corresponds to a different cut.  We can thus write:
\begin{eqnarray}
\label{eq:integral_equation_1}
\mbox{\boldmath ${\cal D}$}_{F}(k) & = & \mbox{\boldmath ${\cal I}$}_{F}(k) + \int\frac{d^{4}p}{(2\pi)^{4}}\; \mbox{\boldmath ${\cal M}$}_{(1)}(k,p)\mbox{\boldmath ${\cal F}$}_{F}(p)\mbox{\boldmath ${\cal D}$}_{F}(p) \nonumber \\
                                            &   & \hspace{0.5in} + \int\frac{d^{4}p}{(2\pi)^{4}}\; \mbox{\boldmath ${\cal M}$}_{(2)}(k,p)\mbox{\boldmath ${\cal F}$}_{B}(p)\mbox{\boldmath ${\cal D}$}_{B}(p) \\
\label{eq:integral_equation_2}
\mbox{\boldmath ${\cal D}$}_{B}(k) & = & \mbox{\boldmath ${\cal I}$}_{B}(k) + \int\frac{d^{4}p}{(2\pi)^{4}}\; \mbox{\boldmath ${\cal M}$}_{(3)}(k,p)\mbox{\boldmath ${\cal F}$}_{F}(p)\mbox{\boldmath ${\cal D}$}_{F}(p) \nonumber \\
                                            &   & \hspace{0.5in} + \int\frac{d^{4}p}{(2\pi)^{4}}\; \mbox{\boldmath ${\cal M}$}_{(4)}(k,p)\mbox{\boldmath ${\cal F}$}_{B}(p)\mbox{\boldmath ${\cal D}$}_{B}(p)
\end{eqnarray}
where we use boldface letters to emphasize the fact that $\mbox{\boldmath ${\cal D}$}$ and $\mbox{\boldmath ${\cal I}$}$ are 4 component column vectors and $\mbox{\boldmath ${\cal K}_{i}$} \equiv \mbox{\boldmath ${\cal M}_{(i)}{\cal F}$}$ are $4 \times 4$ matrices.  See Fig.~\ref{fig:4x4_integral_equation} for the graphical representation of this equation.  To avoid cluttering the equations with indices, we do not write Lorentz indices explicitly (see Eqs.~(\ref{eq:integral_equation_Euclidean_1})-(\ref{eq:integral_equation_Euclidean_2}) for the correct assigment of Lorentz indices).  We again emphasize that, here and in the following, the order of the various fermionic components is not respected; for example, the effective vertex $\mbox{\boldmath ${\cal D}$}_{F}^{\mu}(p)$ has a Dirac structure and should be sandwiched between the two fermionic propagators contained in $\mbox{\boldmath ${\cal F}$}(p)$, something that is not apparent from the present notation.  Only explicit calculations show that the Dirac structure all works out.

The matrix equations~(\ref{eq:integral_equation_1})-(\ref{eq:integral_equation_2}) are amenable to some simplifications.  First note that the side rail matrix ${\cal F}$ can be decomposed into outer products in the following way:
\begin{eqnarray}
\label{eq:Siderail_decomposition}
{\cal F}_{B/F}(p) & = & w_{B/F}(p)u_{B/F}^{T}(p) + h_{B/F}(p)j_{B/F}^{T}(p)
\end{eqnarray}
It can be shown that $h_{B/F}(p)$ and $j_{B/F}(p)$ do not contribute to the final integral equation \cite{Jeon_1995,Gagnon_Jeon_2007}.  The expressions for $u_{B/F}(p)$ and $w_{B/F}(p)$ are:
\begin{eqnarray}
\label{eq:Outer_vectors}
w_{B/F}^{T}(p) & = & \frac{2n_{B/F}(p^{0})\Delta_{B/F}^{+}(p)}{\Gamma_{p\;B/F}}\left( \begin{array}{cccc} 1 & 1 & \frac{(1\pm e^{\beta p^{0}})}{2} & \frac{(1\pm e^{-\beta p^{0}})}{2} \end{array} \right) \nonumber \\
u_{B/F}^{T}(p) & = & \pm\left( \begin{array}{cccc} 1 & 1 & \frac{(1\pm e^{-\beta p^{0}})}{2} & \frac{(1\pm e^{\beta p^{0}})}{2} \end{array} \right)
\end{eqnarray}
Multiplying Eq.~(\ref{eq:integral_equation_1}) by $u_{F}^{T}(k)$ and Eq.~(\ref{eq:integral_equation_2}) by $u_{B}^{T}(k)$ from the LHS, it is possible to reduce the matrix integral equations into one-dimensional integral equations (in ``1-2'' space).  After this operation, the reduced rung kernels $u_{B/F}^{T}(k){\cal M}_{(i)}(k,p)w_{B/F}(p)$ becomes a sum of different cuts.  Using unitarity and KMS relations for 4-point functions in addition to some special relations between the rungs \cite{Jeon_1995,Gagnon_Jeon_2007}, the reduced rung kernel can be further simplified.  After some algebra, we get:
\begin{eqnarray}
\label{eq:Integral_equation_reduced_1}
{\cal D}_{F}(k) & = & -{\cal I}_{F}(k) + \int \frac{d^{4}p}{(2\pi)^{4}} \left[(1+e^{-\beta k^{0}})\left({\cal M}_{22}^{(1)}(k,p)+e^{\beta k^{0}}{\cal M}_{32}^{(1)}(k,p)\right) \bar{\Delta}_{F}^{+}(p)\right] \frac{{\cal D}_{F}(p)}{\Gamma_{p}} \nonumber \\
                &   & \hspace{0.4in} + \int \frac{d^{4}p}{(2\pi)^{4}} \left[(1+e^{-\beta k^{0}})\left({\cal M}_{22}^{(2)}(k,p)+e^{\beta k^{0}}{\cal M}_{32}^{(2)}(k,p)\right) \Delta_{B}^{+}(p)\right] \frac{{\cal D}_{B}(p)}{2E_{p}\gamma_{p}} \\
\label{eq:Integral_equation_reduced_2}
{\cal D}_{B}(k) & = & -{\cal I}_{B}(k) + \int \frac{d^{4}p}{(2\pi)^{4}} \left[(1-e^{-\beta k^{0}})\left({\cal M}_{22}^{(3)}(k,p)+e^{\beta k^{0}}{\cal M}_{32}^{(3)}(k,p)\right) \bar{\Delta}_{F}^{+}(p)\right] \frac{{\cal D}_{F}(p)}{\Gamma_{p}} \nonumber \\
                &   & \hspace{0.4in} + \int \frac{d^{4}p}{(2\pi)^{4}} \left[(1-e^{-\beta k^{0}})\left({\cal M}_{22}^{(4)}(k,p)+e^{\beta k^{0}}{\cal M}_{32}^{(4)}(k,p)\right) \Delta_{B}^{+}(p)\right] \frac{{\cal D}_{B}(p)}{2E_{p}\gamma_{p}}
\end{eqnarray}
where ${\cal D}_{B/F} \equiv \mp u_{B/F}^{T} \mbox{\boldmath ${\cal D}$}_{B/F}$, the ${\cal M}_{ij}$'s ($i,j = 0,...,3$) refer to the matrix components of $\mbox{\boldmath ${\cal M}$}$ (correspond to the different ways of cutting the rung kernel, see Fig.~\ref{fig:4x4_integral_equation}), we defined $\Delta_{F}^{+}(p)\equiv p\!\!\!/\bar{\Delta}_{F}^{+}(p)$ and we used the fact that ${\cal D}_{F} \propto \gamma^{i}$. 

Equations~(\ref{eq:Integral_equation_reduced_1})-(\ref{eq:Integral_equation_reduced_2}), with the non-collinear rungs in Fig.~\ref{fig:Summary_viscosity} and the current insertions given by Eq.~(\ref{eq:Operator_insertion}), contain the necessary ingredients for computing the shear viscosity at leading order (neglecting collinear physics).  One could in principle solve these equations numerically to obtain the effective vertices and then compute the shear viscosity from the Kubo relation~(\ref{eq:Kubo_relations}).

\begin{figure}
\resizebox{4.3in}{!}{\includegraphics{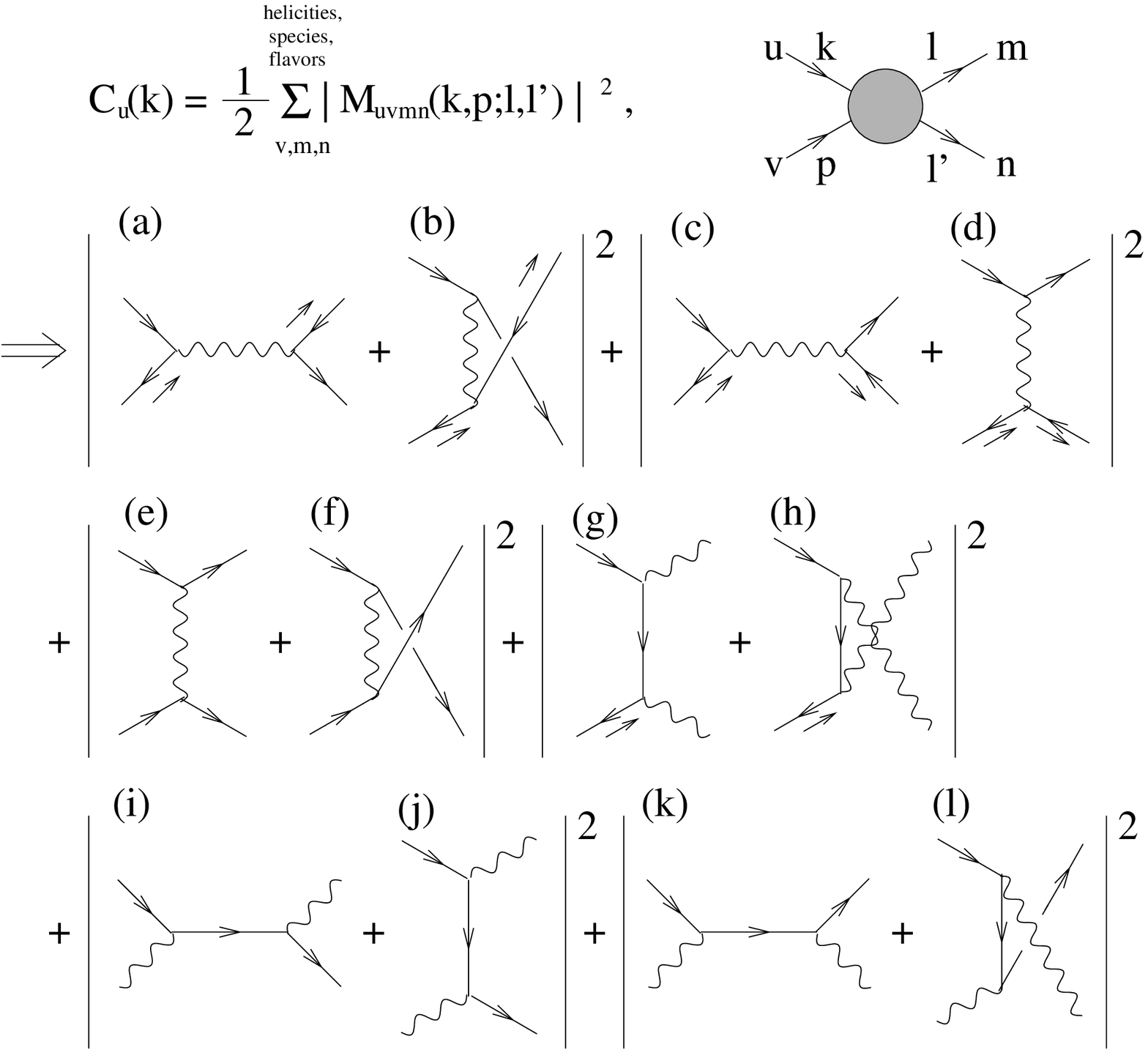}}
\caption{\label{fig:2to2_scatterings_M1_M2} Collision term of the (fermionic) linearized Boltzmann equation including only leading order $2\rightarrow 2$ scatterings \cite{AMY_2003a,AMY_2003b}.  The labels $u$-$n$ represent the species (fermion or photon), flavor and helicity of the excitations (here $u =$ electron).  A straightforward calculation shows the equivalence between the scattering processes and the rungs of Figs.~\ref{fig:2to2_rungs_M1}-\ref{fig:2to2_rungs_M2}.  The correspondence goes as follows (the letters refer to diagrams in Figs.~\ref{fig:2to2_rungs_M1}-\ref{fig:2to2_rungs_M2} and \ref{fig:2to2_scatterings_M1_M2}, respectively): (a1) $=$ (b)$^{2}$ and (d)$^{2}$, (b1) $=$ (e)$^{2}$ and (f)$^{2}$, (c1) $=$ (a)$^{2}$ and (c)$^{2}$, (d1) $=$ (g)$^{2}$ and (h)$^{2}$,(g1) $=$ (g)(h) and (h)(g), (j1) $=$ (e)(f) and (f)(e), (k1) $=$ (b)(a) and (d)(c), (l1) $=$ (a)(b) and (c)(d), (c2) $=$ (j)$^{2}$ and (l)$^{2}$, (d2) $=$ (i)$^{2}$ and (k)$^{2}$, (e2) $=$ (i)(j) and (k)(l), (f2) $=$ (j)(i) and (l)(k).}
\end{figure}

\begin{figure}
\resizebox{4.3in}{!}{\includegraphics{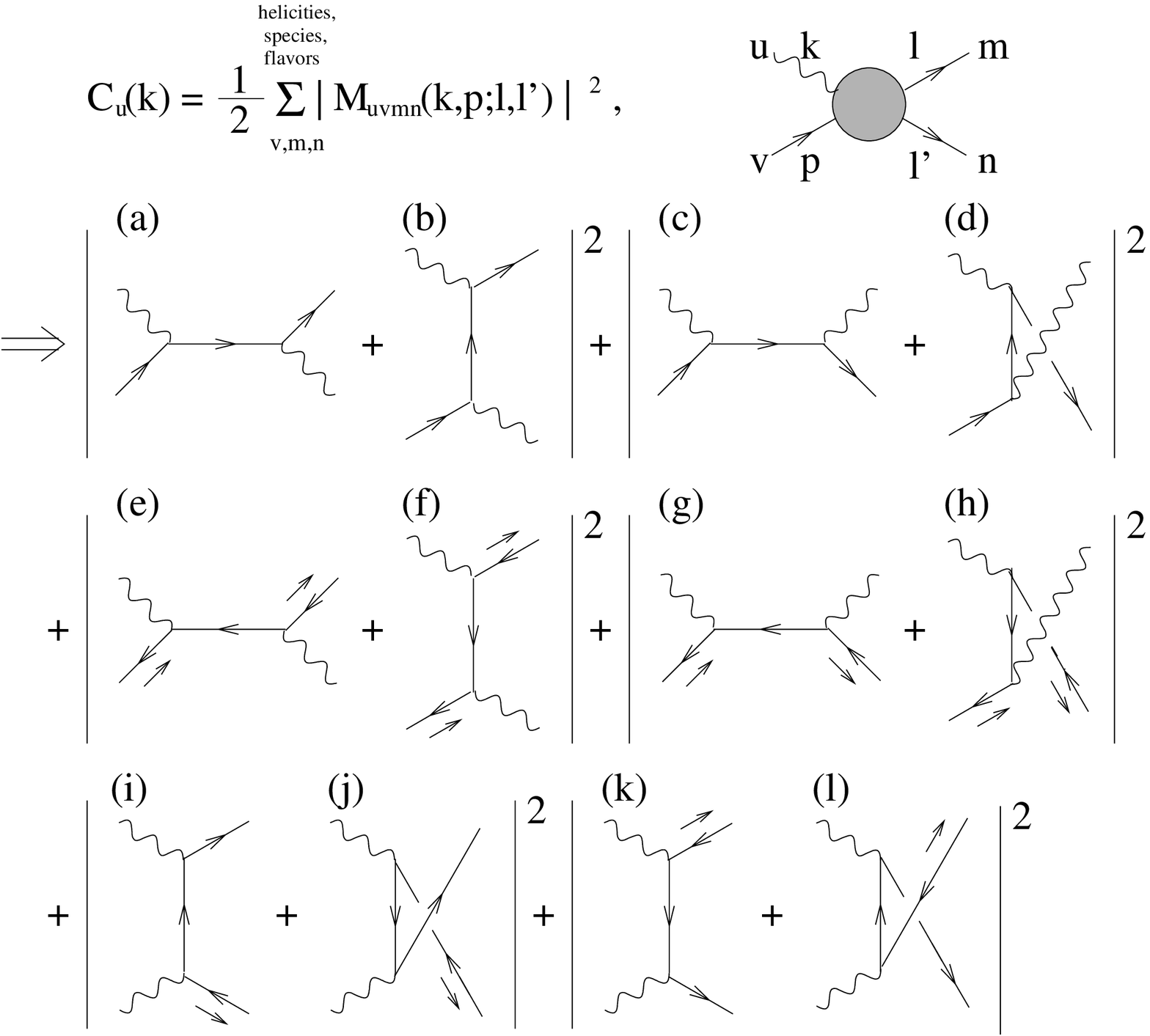}}
\caption{\label{fig:2to2_scatterings_M3_M4} Collision term of the (bosonic) linearized Boltzmann equation including only leading order $2\rightarrow 2$ scatterings \cite{AMY_2003a,AMY_2003b}.  The labels $u$-$n$ represent the species (fermion or photon), flavor and helicity of the excitations (here $u =$ photon).  A straightforward calculation shows the equivalence between the scattering processes and the rungs of Figs.~\ref{fig:2to2_rungs_M3}-\ref{fig:2to2_rungs_M4}.  The correspondence goes as follows (the letters refer to diagrams in Figs.~\ref{fig:2to2_rungs_M3}-\ref{fig:2to2_rungs_M4} and \ref{fig:2to2_scatterings_M3_M4}, respectively):  (a3) $=$ (e)$^{2}$ and (g)$^{2}$,  (d3) $=$ (b)$^{2}$ and (d)$^{2}$, (e3) $=$ (a)$^{2}$ and (c)$^{2}$,  (h3) $=$ (f)$^{2}$ and (h)$^{2}$, (i3) $=$ (e)(f) and (g)(h), (k3) $=$ (b)(a) and (d)(c), (m3) $=$ (a)(b) and (c)(d), (o3) $=$ (f)(e) and (h)(g), (a4) $=$ (i)$^{2}$ and (l)$^{2}$, (b4) $=$ (j)$^{2}$ and (k)$^{2}$, (c4) $=$ (i)(j) and (l)(k), (d4) $=$ (j)(i) and (k)(l).}
\end{figure}

In this paper, we instead want to show the equivalence between Eqs.~(\ref{eq:Integral_equation_reduced_1})-(\ref{eq:Integral_equation_reduced_2}) and the effective kinetic equations of Arnold, Moore and Yaffe \cite{AMY_2003a,AMY_2003b}.  To do that, we first note that the non-collinear rungs in Fig.~\ref{fig:Summary_viscosity} are all made of 4-point functions with two external vertices in the shade and two out of the shade.  Since cut propagators represent nearly on-shell thermal quasi-particles, the non-collinear rungs can be naturally interpreted as $2\rightarrow 2$ scattering processes.  A straightforward calculation shows that the sum of all the leading order non-collinear rungs contained in the kernels ${\cal M}_{(1)}+{\cal M}_{(2)}$ and ${\cal M}_{(3)}+{\cal M}_{(4)}$ can be converted into the square of a scattering matrix, where the scattering processes are given in Figs.~\ref{fig:2to2_scatterings_M1_M2}-\ref{fig:2to2_scatterings_M3_M4}.  It is easy to see that diagrammatically, starting from the scattering processes (see the captions of Figs.~\ref{fig:2to2_scatterings_M1_M2}-\ref{fig:2to2_scatterings_M3_M4} for details).  Multiplying Eq.~\ref{eq:Integral_equation_reduced_1} from the left (right) by the spinor $\bar{u}^{\lambda}(\hat{k})$ ($u^{\lambda}(\hat{k})$), we get: 
\begin{eqnarray}
\label{eq:Integral_equation_scatterings_1}
D_{e}(k) & = & -I_{e}(k) + \int \frac{d^{4}p}{(2\pi)^{4}} \frac{d^{4}l}{(2\pi)^{4}} \frac{d^{4}l'}{(2\pi)^{4}}\; (2\pi)^{4}\delta^{(4)}(k+p-l-l') (1 + e^{-\beta k^{0}}) \nonumber \\
             &   & \hspace{0.5in} \times \left[\frac{1}{2}\sum_{v \in (e,p)\,;\,m,n \in (e,p,\gamma)}^{f,s,h} |M_{evmn}(k,p;l,l')|^{2}\; \bar{\Delta}_{v}^{+}(-p)\bar{\Delta}_{m}^{+}(l)\bar{\Delta}_{n}^{+}(l')\right] \frac{D_{v}(p)}{\Gamma_{p}} \nonumber \\
	     &   & \hspace{0.5in} + \int \frac{d^{4}p}{(2\pi)^{4}} \frac{d^{4}l}{(2\pi)^{4}} \frac{d^{4}l'}{(2\pi)^{4}}\; (2\pi)^{4}\delta^{(4)}(k+p-l-l') (1 + e^{-\beta k^{0}}) \nonumber \\
	     &   & \hspace{0.5in} \times \left[-\frac{1}{2}\sum_{m,n}^{f,s,h} |M_{e\gamma mn}(k,p;l,l')|^{2}\; \bar{\Delta}_{\gamma}^{+}(-p)\bar{\Delta}_{m}^{+}(l)\bar{\Delta}_{n}^{+}(l')\right] \frac{D_{\gamma}(p)}{2E_{p}\gamma_{p}} \\
\label{eq:Integral_equation_scatterings_2}
D_{\gamma}(k) & = & -I_{\gamma}(k) + \int \frac{d^{4}p}{(2\pi)^{4}} \frac{d^{4}l}{(2\pi)^{4}} \frac{d^{4}l'}{(2\pi)^{4}}\; (2\pi)^{4}\delta^{(4)}(k+p-l-l') (1 - e^{-\beta k^{0}}) \nonumber \\
              &   & \hspace{0.5in} \times \left[\frac{1}{2}\sum_{v \in (e,p)\,;\,m,n \in (e,p,\gamma)}^{f,s,h} |M_{\gamma vmn}(k,p;l,l')|^{2}\; \bar{\Delta}_{v}^{+}(-p)\bar{\Delta}_{m}^{+}(l)\bar{\Delta}_{n}^{+}(l')\right] \frac{D_{v}(p)}{\Gamma_{p}} \nonumber \\
	      &   & \hspace{0.5in} + \int \frac{d^{4}p}{(2\pi)^{4}} \frac{d^{4}l}{(2\pi)^{4}} \frac{d^{4}l'}{(2\pi)^{4}}\; (2\pi)^{4}\delta^{(4)}(k+p-l-l') (1 - e^{-\beta k^{0}}) \nonumber \\
	      &   & \hspace{0.5in} \times \left[-\frac{1}{2}\sum_{m,n}^{f,s,h} |M_{\gamma\gamma mn}(k,p;l,l')|^{2}\; \bar{\Delta}_{\gamma}^{+}(-p)\bar{\Delta}_{m}^{+}(l)\bar{\Delta}_{n}^{+}(l')\right] \frac{D_{\gamma}(p)}{2E_{p}\gamma_{p}}
\end{eqnarray}
where the $M_{uvmn}(k,p;l,l')$'s are the $2\rightarrow 2$ scattering processes shown in Figs.~\ref{fig:2to2_scatterings_M1_M2}-\ref{fig:2to2_scatterings_M3_M4}, we have used explicit labels for excitations ($e = $electron, $p = $positron, $\gamma = $photon), we have defined $D_{F}(k) \equiv \bar{u}^{\lambda}(\hat{k}){\cal D}_{F}(k)u^{\lambda}(\hat{k})$ and redefined $D_{B}(k) \equiv {\cal D}_{B}(k)$.  The sums are over flavors (f), species (s) and helicities (h).

We also note that the integrals in Eqs.~(\ref{eq:Integral_equation_reduced_1})-(\ref{eq:Integral_equation_reduced_2}) are over $d^{4}p$, while kinetic equations deal with on-shell quasi-particle propagation.  Thus to make the connection between the two approaches clearer, we use the delta functions present in the cut propagators to put excitations on-shell in Eqs.~(\ref{eq:Integral_equation_reduced_1})-(\ref{eq:Integral_equation_reduced_2}).  We remark here that at finite temperature, both positive and negative energies appear in the cut lines.  In addition, the distribution functions inside the cut propagators provide the thermal phase space for the gain and loss terms in the kinetic equations.

Defining the deviations from equilibrium $\chi_{B}(k) \equiv  D_{B}(k)/2E_{k}\gamma_{k}$ and $\chi_{F}(k) \equiv D_{F}(k)/\Gamma_{k}$ and noting that $\gamma_{k}$ and $\Gamma_{k}$ can be written as squares of scattering processes, we follow the same procedure as in Ref.~\cite{Gagnon_Jeon_2007} to get:
\begin{eqnarray}
\label{eq:Integral_equation_AMY_1}
-(1-n_{e}(k^{0}))I_{e}(k) & = & \frac{1}{2}\int \frac{d^{3}p}{(2\pi)^{3}2E_{p}} \frac{d^{3}l}{(2\pi)^{3}2E_{l}} \frac{d^{3}l'}{(2\pi)^{3}2E_{l'}}\;  (2\pi)^{4}\delta^{(4)}(k+p-l-l')\;  \nonumber \\
            &   & \times \sum_{v,m,n}^{f,s,h} |M_{evmn}(k,p;l,l')|^{2}\; n_{e}(k^{0})n_{v}(p^{0})(1\pm n_{m}(l^{0}))(1\pm n_{n}({l'}^{0})) \nonumber \\
	    &   & \times \left[\chi_{e}(k) + \chi_{v}(p) - \chi_{m}(l) - \chi_{n}(l')\right] \\
\label{eq:Integral_equation_AMY_2}
-(1+n_{\gamma}(k^{0}))I_{\gamma}(k) & = & \frac{1}{2}\int \frac{d^{3}p}{(2\pi)^{3}2E_{p}} \frac{d^{3}l}{(2\pi)^{3}2E_{l}} \frac{d^{3}l'}{(2\pi)^{3}2E_{l'}}\;  (2\pi)^{4}\delta^{(4)}(k+p-l-l')\;  \nonumber \\
            &   & \times \sum_{v,m,n}^{f,s,h} |M_{\gamma vmn}(k,p;l,l')|^{2}\; n_{\gamma}(k^{0})n_{v}(p^{0})(1\pm n_{m}(l^{0}))(1\pm n_{n}({l'}^{0})) \nonumber \\
	    &   & \times \left[\chi_{\gamma}(k) + \chi_{v}(p) - \chi_{m}(l) - \chi_{n}(l')\right]
\end{eqnarray}
These coupled integral equations are identical to the ones obtained by Arnold, Moore and Yaffe \cite{AMY_2003b,AMY_2003a} using effective kinetic theory (without collinear processes).


\subsection{Integral equation with collinear singularities (${\cal N}$)}
\label{sec:Integral_equation_with_collinear}

The case with collinear rungs can be treated using the same tools used for non-collinear rungs.  Since the steps are essentially similar, we focus in this section on how to write the collinear version of Eqs.~(\ref{eq:Integral_equation_scatterings_1})-(\ref{eq:Integral_equation_scatterings_2}) and how to properly write the leading order collinear rungs as $1+N\rightarrow 2+N$ scattering processes.  

\begin{figure}
\resizebox{5.5in}{!}{\includegraphics{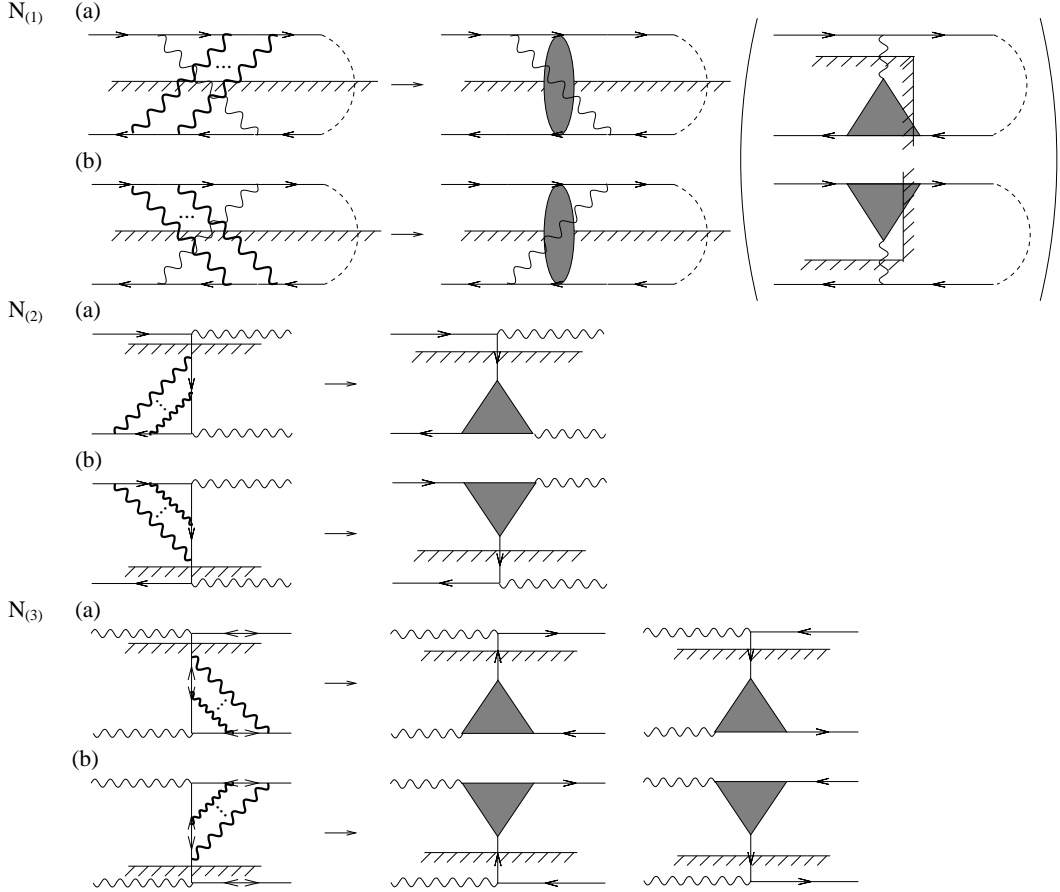}}
\caption{\label{fig:1to2_rungs_resummed} Rungs that are part of the leading order collinear analysis.  A thick line corresponds to a soft propagator.  Due to collinear singularities, adding a soft photon is an $O(1)$ correction; thus for each rung kernel ${\cal N}_{(i)}$, an infinite number of rung types contribute at leading order.  These soft vertex corrections can be resummed using another integral equation (c.f. Eq.~(\ref{eq:Vertex_resummation})).  For ${\cal N}_{(1)}$, it is not readily apparent that these soft photons correspond to vertex corrections.  To see that, notice that the rung is effectively closed with a cut propagator coming from the reduction of the side rail matrix ${\cal F}$ (c.f. Eqs.~(\ref{eq:Siderail_decomposition})-(\ref{eq:Outer_vectors})); using this fact, we see that the resummed ${\cal N}_{(1)}$ rung is equivalent to the one in parenthesis.}
\end{figure}

According to the power counting of Sect.~\ref{sec:Power_counting_with_collinear}, the rung kernels are given by ${\cal N}_{(1)} = (u1)+(x1)$, ${\cal N}_{(2)} = (u2)+(v2)$ and ${\cal N}_{(3)} = (u3)+(v3)+(w3)+(x3)$, along with all similar rungs with an arbitrary number of soft photon corrections (see Fig.~\ref{fig:1to2_rungs_resummed}) \footnote{Note that rungs (v1), (w1) were included in \cite{Gagnon_Jeon_2007} instead of rungs (u1), (x1).  As explained in Sect.~\ref{sec:Power_counting_with_collinear}, the size of both sets of rungs is (naively) $O(e^{4})$, but (v1),(w1) turn out to be subleading due to a cancellation.  However, since both sets of rungs can be seen as vertex corrections, the final result is the same \cite{Gagnon_Jeon_2007_erratum}.}.  For ${\cal N}_{(2)}$ and ${\cal N}_{(3)}$, this infinite number of soft photons corresponds to vertex corrections.  This is also the case for ${\cal N}_{(1)}$, although it is less apparent; the key point to understand this is to note that the reduction of the side rail matrix ${\cal F}$ into a cut propagator in the pinch limit (c.f. Eqs.~(\ref{eq:Siderail_decomposition})-(\ref{eq:Outer_vectors})) effectively closes the rung, making the soft exchange photon between the two side rails look like a vertex correction (see Fig.~\ref{fig:1to2_rungs_resummed}).

\begin{figure}
\resizebox{4.3in}{!}{\includegraphics{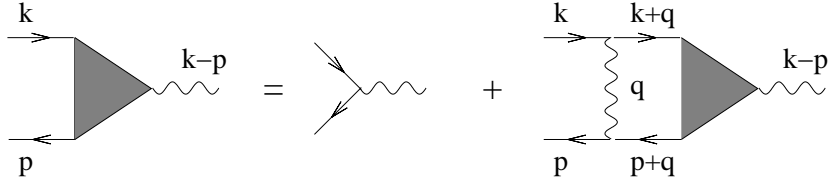}}
\caption{\label{fig:Integral_equation_collinear} Illustration of the integral equation~(\ref{eq:Vertex_resummation}).  The kernel ${\cal N}_{\rm coll}$ is made of only a single soft photon exchange.}
\end{figure}

This infinite number of vertex corrections due to collinear singularities can be resummed by defining an effective vertex ${\cal V}_{F}(k,p)$ that satisfies the following integral equation:
\begin{eqnarray}
\label{eq:Vertex_resummation}
{\cal V}_{F}(k,p) & = & {\cal I}_{F}(k,p) + \int\frac{d^{4}q}{(2\pi)^{4}} {\cal N}_{\rm coll}(k,p,q){\cal F}_{F}(k,p,q){\cal V}_{F}(p,q)
\end{eqnarray}
where ${\cal N}_{\rm coll}$ is a rung with a single soft photon exchange and the external photon is collinear with the electron.  See Fig.~\ref{fig:Integral_equation_collinear} for an illustration of this integral equation.  This resummation is relevant for photon production including the LPM effect and is done in great detail in \cite{AMY_2001a,AMY_2001b}.  Instead of using the closed-time-path or ``1-2'' formalism, they use the r/a formalism and are able to put the integral equation in a form convenient for numerical purposes.

Applying the reduction procedure of Sect.~\ref{sec:Integral_equation_without_collinear} to the collinear rung kernels of Fig.~\ref{fig:1to2_rungs_resummed}, it is possible to write a set of integral equations similar to Eqs.~(\ref{eq:Integral_equation_reduced_1})-(\ref{eq:Integral_equation_reduced_2}) (with the ${\cal M}$'s replaced by ${\cal N}$'s).  Note that ${\cal N}_{32}^{(2)} = {\cal N}_{32}^{(3)} = 0$ (that particular cut does not exist for 3-point functions) and ${\cal N}_{32}^{(1)}$ is subleading; there is thus no ${\cal N}_{32}$ contribution in the collinear version of Eqs.~(\ref{eq:Integral_equation_reduced_1})-(\ref{eq:Integral_equation_reduced_2}).

\begin{figure}
\resizebox{\textwidth}{!}{\includegraphics{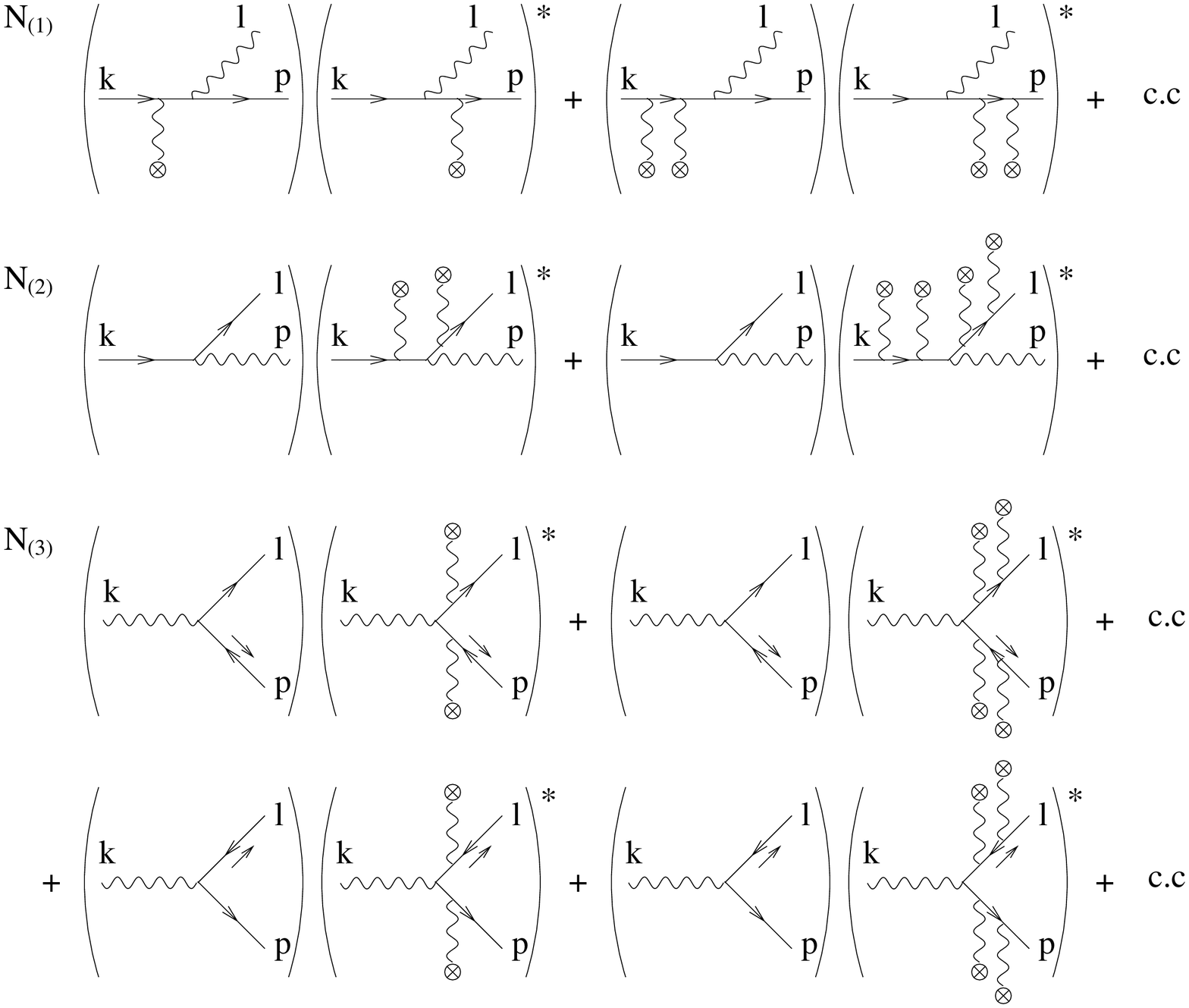}}
\caption{\label{fig:1to2_scatterings} $1+N \rightarrow 2+N$ collinear scatterings included in Eqs.~(\ref{eq:Integral_equation_collinear_1})-(\ref{eq:Integral_equation_collinear_2}).  Scatterings with soft photons are represented by a wiggly line ending with a crossed circle.  These collinear processes correspond to the ones identified in Ref.~\cite{AMY_2002} as responsible for the LPM effect.}
\end{figure}

To obtain the collinear version of Eqs.~(\ref{eq:Integral_equation_scatterings_1})-(\ref{eq:Integral_equation_scatterings_2}), we need to express the collinear rungs in Fig.~\ref{fig:1to2_rungs_resummed} in terms of $1\rightarrow 2$ collinear scattering processes.  We first note that all the collinear rung kernels ${\cal N}^{(i)}$ can be written as the multiplication of a bare $1\rightarrow 2$ vertex $V_{1\rightarrow 2}^{\mu}$ and a resummed vertex ${\cal V}_{F}^{\nu}$.  For instance, we have ${\cal N}_{22\;(a)}^{(1)} = \left[-iV_{1\rightarrow 2}^{\mu}\right]\Delta_{B\;\mu\nu}^{+}(k-p)\left[-i{\cal V}_{F}^{\nu}\right]^{*}$ and ${\cal N}_{22\;(b)}^{(1)} =   \left[-i{\cal V}_{F}^{\mu}\right]\Delta_{B\;\mu\nu}^{+}(k-p)\left[-iV_{1\rightarrow 2}^{\nu}\right]^{*}$.  Note also that ${\cal N}_{22\;(a)}^{(1)}$ and ${\cal N}_{22\;(b)}^{(1)}$ are complex conjugate of each other; this is the consequence of the trace over the fermion structure in the Kubo formula, which allows us in this case to freely interchange the Dirac structure.  The sum of ${\cal N}_{22\;(a)}^{(1)}$ and ${\cal N}_{22\;(b)}^{(1)}$ is thus twice the real part of ${\cal N}_{22\;(a)}^{(1)}$.  The other rung kernels ${\cal N}^{(2)}$ and ${\cal N}^{(3)}$ can be similarly written.  Following the procedure in Sect.~\ref{sec:Integral_equation_without_collinear}, we obtain:
\begin{eqnarray}
\label{eq:Integral_equation_collinear_1}
D_{e}(k) & = & -I_{e}(k) + \int \frac{d^{4}p}{(2\pi)^{4}} \frac{d^{4}l}{(2\pi)^{4}}\; (2\pi)^{4}\delta^{(4)}(k-p-l) (1+e^{-\beta k^{0}}) \nonumber \\
            &   & \hspace{0.7in} \times \left[2\mbox{Re}[{\cal V}_{F}^{\mu}V_{1\rightarrow 2\;\mu}^{*}]_{e\gamma}^{e}(k;p,l)\; \bar{\Delta}_{e}^{+}(p)\Delta_{\gamma}^{+}(l)\right] \frac{D_{e}(p)}{\Gamma_{p}} \nonumber \\
            &   & \hspace{0.6in} + \int \frac{d^{4}p}{(2\pi)^{4}} \frac{d^{4}l}{(2\pi)^{4}}\; (2\pi)^{4}\delta^{(4)}(k-p-l) (1+e^{-\beta k^{0}}) \nonumber \\
            &   & \hspace{0.7in} \times \left[2\mbox{Re}[{\cal V}_{F}^{\mu}V_{1\rightarrow 2\;\mu}^{*}]_{\gamma e}^{e}(k;p,l)\; \Delta_{\gamma}^{+}(p)\bar{\Delta}_{e}^{+}(l)\right] \frac{D_{\gamma}(p)}{2E_{p}\gamma_{p}} \\
\label{eq:Integral_equation_collinear_2}
D_{\gamma}(k) & = & -I_{\gamma}(k) + \int \frac{d^{4}p}{(2\pi)^{4}} \frac{d^{4}l}{(2\pi)^{4}}\; (2\pi)^{4}\delta^{(4)}(k-p-l) (1-e^{-\beta k^{0}}) \nonumber \\
            &   & \hspace{0.7in} \times \left[\sum_{v,m \in (e,p)}^{f,s,h} 2\mbox{Re}[{\cal V}_{F}^{\mu}V_{1\rightarrow 2\;\mu}^{*}]_{vm}^{\gamma}(k;p,l)\; \bar{\Delta}_{v}^{+}(p)\bar{\Delta}_{m}^{+}(l)\right] \frac{D_{v}(p)}{\Gamma_{p}}
\end{eqnarray}
Figure~\ref{fig:1to2_scatterings} shows the $1\rightarrow 2$ collinear scattering processes corresponding to the rungs in Fig.~\ref{fig:1to2_rungs_resummed}.  These scattering processes are the ones that are responsible for the LPM effect \cite{AMY_2002}.  The equivalence with kinetic theory can again be seen by defining the deviations from equilibrium $\chi_{B}(k) \equiv D_{B}(k)/2E_{p}\gamma_{k}$ and $\chi_{F}(k) \equiv D_{F}(k)/\Gamma_{k}$ and following the procedure in Sect.~\ref{sec:Integral_equation_without_collinear}.  The final result is:
\begin{eqnarray}
\label{eq:Integral_equation_AMY_LPM_1}
-(1-n_{e}(k^{0}))I_{e}(k) & = & \frac{1}{2}\int \frac{d^{3}p}{(2\pi)^{3}2E_{p}} \frac{d^{3}l}{(2\pi)^{3}2E_{l}}\;  (2\pi)^{4}\delta^{(4)}(k-p-l)\; \sum_{v,m}^{f,s,h} 2\mbox{Re}[{\cal V}_{F}^{\mu}V_{1\rightarrow 2\;\mu}^{*}]_{vm}^{e}(k;p,l) \nonumber \\
           &   & \times \left[ n_{e}(k^{0})(1\pm n_{v}(p^{0}))(1\pm n_{m}(l^{0}))\left(\chi_{e}(k) - \chi_{v}(p) - \chi_{m}(l)\right)\right] \nonumber \\
	         &   & + \int \frac{d^{3}p}{(2\pi)^{3}2E_{p}} \frac{d^{3}l}{(2\pi)^{3}2E_{l}}\;  (2\pi)^{4}\delta^{(4)}(k+p-l)\; \sum_{v,m}^{f,s,h} 2\mbox{Re}[{\cal V}_{F}^{\mu}V_{1\rightarrow 2\;\mu}^{*}]_{m}^{ev}(k,p;l) \nonumber \\
           &   & \times \left[ n_{e}(k^{0})n_{v}(p^{0})(1\pm n_{m}(l^{0}))\left(\chi_{e}(k) + \chi_{v}(p) - \chi_{m}(l)\right)\right] \\
\label{eq:Integral_equation_AMY_LPM_2}
-(1+n_{\gamma}(k^{0}))I_{\gamma}(k) & = & \frac{1}{2}\int \frac{d^{3}p}{(2\pi)^{3}2E_{p}} \frac{d^{3}l}{(2\pi)^{3}2E_{l}}\;  (2\pi)^{4}\delta^{(4)}(k-p-l)\; \sum_{v,m}^{f,s,h} 2\mbox{Re}[{\cal V}_{F}^{\mu}V_{1\rightarrow 2\;\mu}^{*}]_{vm}^{\gamma}(k;p,l) \nonumber \\
           &   & \times \left[ n_{\gamma}(k^{0})(1\pm n_{v}(p^{0}))(1\pm n_{m}(l^{0}))\left(\chi_{\gamma}(k) - \chi_{v}(p) - \chi_{m}(l)\right)\right] \nonumber \\
	         &   & + \int \frac{d^{3}p}{(2\pi)^{3}2E_{p}} \frac{d^{3}l}{(2\pi)^{3}2E_{l}}\;  (2\pi)^{4}\delta^{(4)}(k+p-l)\; \sum_{v,m}^{f,s,h} 2\mbox{Re}[{\cal V}_{F}^{\mu}V_{1\rightarrow 2\;\mu}^{*}]_{m}^{\gamma v}(k,p;l) \nonumber \\
           &   & \times \left[ n_{\gamma}(k^{0})n_{v}(p^{0})(1\pm n_{m}(l^{0}))\left(\chi_{\gamma}(k) + \chi_{v}(p) - \chi_{m}(l)\right)\right]
\end{eqnarray}
This last equation is identical to the linearized Boltzmann equation with the transverse momenta non-integrated of Arnold, Moore and Yaffe \cite{AMY_2003b,AMY_2003a} obtained using kinetic theory.

\section{Conclusion}
\label{sec:Conclusion}

In this paper, we have derived the integral equations needed for the calculation of shear viscosity at leading order in hot QED.  A visual summary of our calculation is presented in Fig.~\ref{fig:Summary_viscosity}.  Our calculation includes all the necessary $2\rightarrow 2$ scatterings and $1\rightarrow 2$ collinear scatterings responsible for the LPM effect.  The equivalence with the effective kinetic theory results of Arnold, Moore and Yaffe has also been established.

We used Ward-like identities to put constraints on the form of the rung kernels.  In the case of electrical conductivity, these constraints are directly related to gauge symmetry and gauge invariance is explicitly enforced.  The issue is not so clear in the case of shear viscosity, since the Ward-like identities used to obtain the constraints derive from space-time symmetries.  But since the constraints~(\ref{eq:Constraint_M_1})-(\ref{eq:Constraint_M_4}) have the same form as those obtained from the usual Ward identity in the case of electrical conductivity, then Eqs.~(\ref{eq:Constraint_M_1})-(\ref{eq:Constraint_M_4}) should be sufficient to preserve gauge invariance; we have shown explicitly that this is indeed the case.

The formalism developped in the present paper could in principle be applied to other transport coefficients, such as the bulk viscosity.  In scalar theories, bulk viscosity is sensitive to number changing processes at leading order whereas shear viscosity is not, making the calculation of the former more involved.  In gauge theories, all transport coefficients are sensitive to these processes even at leading order.  Thus only minor modifications (e.g. changing the current insertion) are required to adapt the present formalism to the computation of other transport coefficients in QED.

Other extensions include the generalization of our formalism to non-abelian gauge theories and the computation of transport coefficients beyond leading order.  Of course, generalizing our calculation to say QCD would be quite demanding, because of the presence of ghosts and a more complicated collinear resummation (due to the self-interaction of gluons among themselves).  Going beyond leading order would also be very challenging, if just for the very subtle power counting that would be required or the possibility of ``new physics'' (e.g. new singularities that would require additional resummations) appearing in the calculations.

\appendix

\section{Gauge invariance of shear viscosity}
\label{app:Gauge_invariance}

In this appendix, we present arguments showing the independence of the shear viscosity on the gauge parameter $\xi$ at leading order in QED.  The starting point is the Kubo relation~(\ref{eq:Kubo_relation_schematic}) with the accompanying integral equations~(\ref{eq:integral_equation_schematic_1})-(\ref{eq:integral_equation_schematic_2}).  Since the current insertions ${\cal I}_{B/F}$ do not depend on $\xi$, the goal is to show that the side rails ${\cal F}_{B/F}$ and the effective vertices ${\cal D}_{B/F}$ do not have any gauge parameter dependence.  The side rails ${\cal F}_{F}$ may depend on $\xi$ through their resummed self-energies (we come back to ${\cal F}_{B}$ later).  The effective vertices are solution to the integral equations~(\ref{eq:integral_equation_schematic_1})-(\ref{eq:integral_equation_schematic_2}); ${\cal D}_{B/F}$ may thus depend on $\xi$ through the side rails ${\cal F}_{B/F}$ and the four rung kernels ${\cal M}$.  Since the Ward-like identity constraints~(\ref{eq:Constraint_M_1})-(\ref{eq:Constraint_M_4}) establish a direct relation between the rungs and the self-energies resummed into the side rail propagators, the only thing we have to show is that the resummed self-energies are gauge parameter independent.

Before presenting our arguments, note that there are two different types of Ward identities in this study.  The first one is $k_{\mu}A^{\mu} = 0$ and is discussed in e.g. \cite{Peskin_Schroeder_1995}.  The second one is $k_{\mu}{\cal D}^{\mu} \propto \Sigma_{I}(k)$ and is discussed in Sect.~\ref{sec:Ward_identity}.  The latter is valid only in the kinematic limit relevant for transport coefficients and when the two external resummed propagators pinch.

For definiteness, let's consider fermionic self-energies (bosonic self-energies can be analyzed in a similar way).  The leading order fermionic self-energies included in the side rail propagators are shown in Fig.~\ref{fig:Summary_viscosity}.  They can be divided into two subgroups, those containing and those not containing collinear singularities.  Using the usual Ward identity $k_{\mu}A^{\mu} = 0$ (where $A^{\mu}$ is any amplitude with all external excitations on-shell), we can show that all the non-collinear self-energies are gauge parameter independent.  For example, the first fermionic self-energy in Fig.~\ref{fig:Summary_viscosity} contains two uncut (i.e. off-shell) photons.  The $\xi$ dependent part in the photon propagator is proportional to $\xi k^{\mu}k^{\nu}$ and we can send each of those $k$'s to a vertex.  Since all the electrons are cut (i.e. on-shell), the four vertices dotted with a $k$ are zero due to the Ward identity.  The first self-energy is thus $\xi$ independent.  The second fermionic self-energy in Fig.~\ref{fig:Summary_viscosity} contains two cut photons that correspond to external (thermal) excitations.  From the Ward identity, any amplitude dotted with a $k^{\mu}$ is zero; thus the $\xi k^{\mu}k^{\nu}$ part of the cut propagators does not contribute.  Using similar arguments, we can show that the other two non-collinear fermionic self-energies are also gauge parameter independent.  

The case of collinear fermionic self-energies is slightly more involved than the non-collinear one, since self-energies with an infinite number of soft corrections to the hard photon vertex contribute at leading order.  To deal with this infinite number of soft corrections, note that all the electron propagators are nearly on-shell (i.e. on-shell to within $O(e^{2}T^{2})$), since the hard photon is collinear to the incoming electron and the soft photons can only push it out of his mass shell with $O(eT)$ kicks.  This implies that the electron wavefunction obeys a ``nearly on-shell Dirac equation'' $(p\!\!\!/-m)u(p) \sim O(eT)$.  From this nearly on-shell Dirac equation, we deduce that the Ward identity obeyed by each vertex with two incoming nearly on-shell electrons is $k_{\mu}A^{\mu} \sim O(eT)$.  As for the non-collinear case, we can use this ``nearly on-shell'' Ward identity to deal with the $\xi k^{\mu}k^{\nu}$ term in each photon.  Since we must use the nearly on-shell Ward identity twice for each photon, the $\xi$ dependent part of the self-energy is suppressed by at least $O(g^{2})$ compared to the rest of the diagram.  Thus the gauge non-invariance of collinear fermionic self-energies only appears at higher orders.

Another potential source of $\xi$ dependence is ${\cal F}_{B}$ (i.e. the bosonic side rail propagators).  To see that ${\cal F}_{B}$ does not give rise to any $\xi$ dependence, note that two pinching propagators can always be expressed as a cut propagator (c.f. Eq.~(\ref{eq:Outer_vectors})).  Consequently, the $\xi k^{\mu}k^{\nu}$ term in ${\cal F}_{B}$ is always on the external leg of an amplitude and vanish by the Ward identity.  This completes our argumentation and shows that the shear viscosity is gauge invariant at leading order.

\end{document}